\documentclass[aps, pra, 
			   final, reprint, 
			   twocolumn]{revtex4-1}

\usepackage{amsmath}
\usepackage{amsfonts}
\usepackage{amssymb}
\usepackage{gensymb}
\usepackage{graphicx}
\usepackage[utf8]{inputenc}
\usepackage[colorlinks=true, linkcolor=black, urlcolor=blue, citecolor=black, anchorcolor=black]{hyperref}

\usepackage{enumitem}
\setlist{nosep}

\def\SItxt{See Supplemental Material at [URL will be inserted by publisher] for 
	Materials and Methods,
	a unified perturbation theory of the OSE for weak field and harmonic probes,
	additional numerical simulations of pump-THG-probe,
	comparison of pump-SHG-probe and our OSE theory,
	demonstration of probe-induced OSE, 
	and additional pump-THG-probe measurements on various WS\textsubscript{2} morphologies.
}

\begin{document}
\title{Quantum interference between the optical Stark effect and resonant harmonic generation in WS\textsubscript{2}}
\author{Darien J. Morrow}
\author{Daniel D. Kohler}
\author{Yuzhou Zhao}
\author{Jason M. Scheeler}
\author{Song Jin}
\author{John C. Wright}
	\email{wright@chem.wisc.edu}
\affiliation{Department of Chemistry,
	University of Wisconsin--Madison,
	1101  Ave, 
	Madison, WI 53706, United States}

%\date{\today}

\begin{abstract}
	An applied field can modulate optical signals by resonance shifting via the Stark effect. 
	The optical Stark effect (OSE) uses ultrafast light in the transparency region of a material to shift resonances with speeds limited by the pulse duration or system coherence.
	In this Letter we investigate the OSE in resonant optical harmonic generation (OHG) using the ground state exciton transition of WS\textsubscript{2} with a variety of morphologies. 
	Multidimensional pump-harmonic-probe measurements, in which the probe is second- or third-harmonic emission, reveal not only large Stark shifts that are commensurate with the large optical susceptibilities common to WS\textsubscript{2} excitons, but also behaviors more complex than simple OSE treatments predict.
	We show how a new manifestation of the Stark Effect, brought forth by coherent photon exchange between the pump and OHG fundamental fields, can strongly enhance or suppress OHG.
	%This new interference effect is promising for ultrafast modulation applications and important for the emerging field of pump-non-linear-probe spectroscopy.
\end{abstract}

\maketitle

% --- introduction ---------------------------------------------------------------------------------

\begin{figure}[!htbp]
	\centering
	\includegraphics[trim={0cm 1cm 0cm 1.5cm},clip, width=\linewidth]{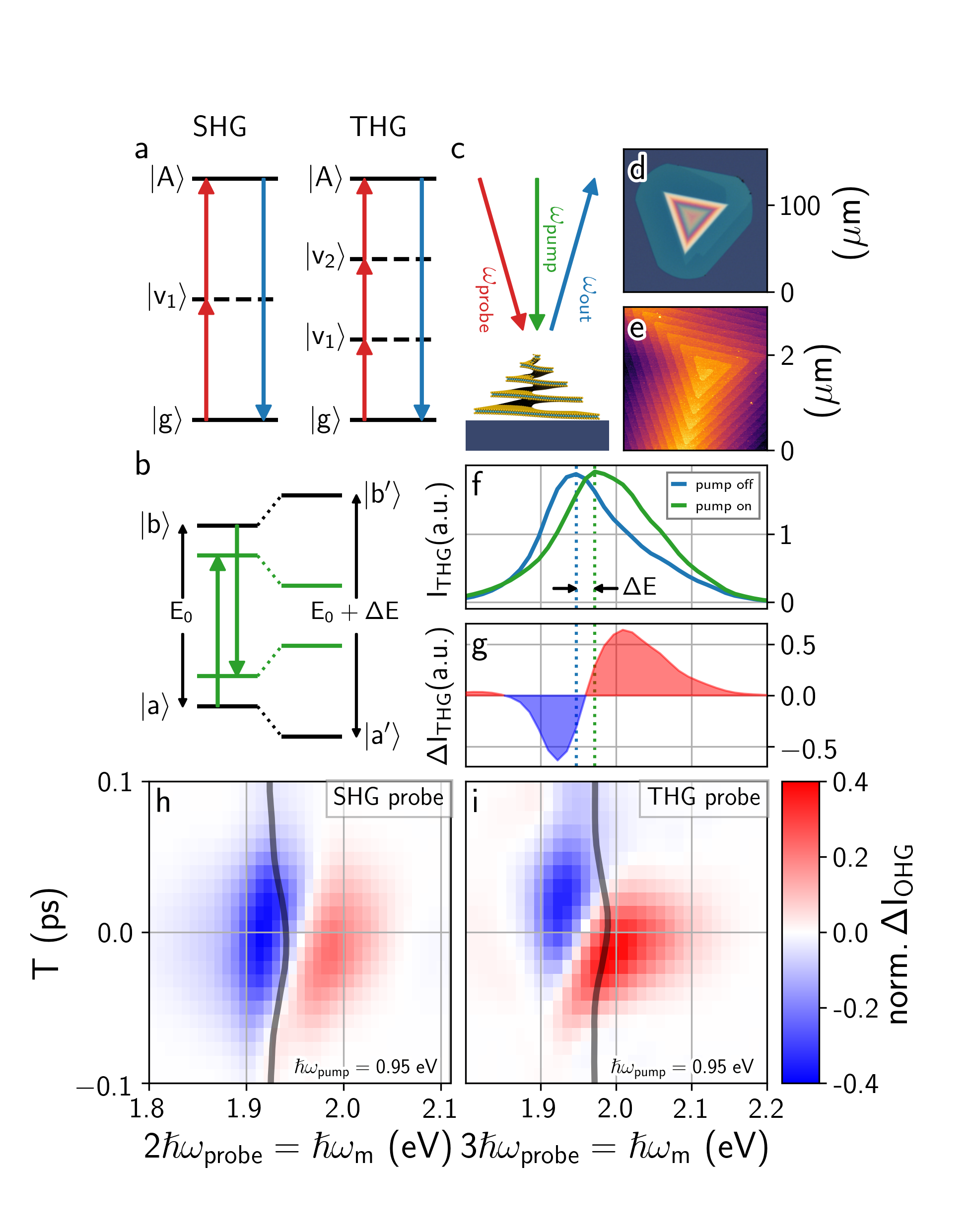}
	\caption{
		Overview of the optical Stark effect and optical harmonic generation.
		(a) Energy level diagrams showing how a probe (red) creates an A exciton coherence which then emits at a frequency (blue) which is a harmonic of the probe.
		(b) Non-resonant optical Stark effect in which a pump (green) drives the $a	\leftrightarrow b$ transition which results in photon-dressed states with energy $E_0 + \Delta E$.
		(c) Illustration of the experimental geometry with $\omega_{\text{out}} = n\cdot\omega_{\text{probe}}$ and $n$ being 2 for SHG and 3 for THG.
		(d, e) Optical and atomic force microscopy images of the WS\textsubscript{2} screw-dislocation pyramid on Si/SiO\textsubscript{2} studied in the main text. 
		(f) THG spectrum of WS\textsubscript{2} pyramid with NIR pump on and off.
		(g) Difference between THG spectra in (f). 
		(h,i) Difference between unpumped and pumped OHG spectrum (h: SHG, i:THG) for different pump-probe time delays, $T$.
		Both plots share the colormap with red (blue) being an increase (decrease) in harmonic generation upon pump excitation.
		The thick black lines are the center-of-mass of the pumped harmonic generation spectrum.
		The pump frequency is $\hbar\omega=0.95$ eV.
	}\label{fig:intro} 
\end{figure}

Optical harmonic generation (OHG) is an important light generation mechanism and a ubiquitous probe in microscopic analysis.
OHG occurs when a strong light field, $E$, of frequency $\omega$, drives a non-linear polarization that coherently radiates new light fields at the harmonics of the original frequency, $\{2\omega, 3\omega, \dots\}$ (\autoref{fig:intro}a).\cite{Franken_Weinreich_1961, Maker_Savage_1962}
OHG is sensitive to phase-matching,\cite{Maker_Savage_1962} crystal orientation,\cite{Li_Heinz_2013} thickness, applied DC electric and magnetic fields,\cite{Seyler_Xu_2015, Sirtori_Cho_1992, Lafrentz_Bayer_2013, Brunne_Bayer_2015} and having frequency components resonant with transitions.\cite{Wang_Urbaszek_2015, Wang_Davis_1975}
OHG spectroscopy, in which the excitation frequency is changed, is a selective probe of semiconductor materials and is sensitive to transitions in ways complementary to that of traditional absorption and reflection probes.\cite{Morrow_Wright_2018, Morrow_Wright_2019,Lafrentz_Bayer_2013}

A related optical process, also requiring strong light fields below the band edge, is the optical Stark effect (OSE).
In the OSE, a non-resonant light field (the pump) coherently drives a system, creating photon-dressed states which hybridize with the system's original eigenstates, shifting their energies (\autoref{fig:intro}b).\cite{Autler_Townes_1955, Bakos_1977, Sussman_2011, Boyd_2008}
The change in transition energy due to the pump is
\begin{align}
	\Delta E = \frac{\left|\mu_{ab}\right|^2 \mathcal{E}^2}{E_0-\hbar\omega_{\text{pump}}}, \label{eq:starkshift}
\end{align}
in which $\mathcal{E}$ is the pump field amplitude,
$\mu_{ab}$ is the transition dipole between states $a$ and $b$,
and $E_0 \equiv E_b - E_a$ is the unpumped transition energy.
When the pump is detuned below the transition resonance, the hybridized states repel and the OSE manifests as a blue-shift of the resonance, which adiabatically follows the pump's envelope.
The OSE is well-known in semiconductor exciton systems, but it is typically observed via a weak electric field probe.\cite{Mysyrowicz_1986,Peyghambarian_Joffre_1989,Knox_SchmittRink_1989,Binder_Schafer_1990}
% DDK: an optional plug; this probably isn't necessary to include, though we have citations to back it up
%The OSE has potential applications in photon chemistry, coherent control, and quantum computing.\textcolor{red}{CITE}

The OSE can have an important interplay with OHG.
In resonant OHG, the OSE alters resonance enhancement, which can modulate OHG efficiency.\cite{Holthaus_Hone_1994,Elgin_New_1976}
Since the OSE is adiabatic, its ultrafast control of OHG may suit photonics applications like optical modulators.\cite{Sun_Wang_2016}
The OSE also alters the free induction decay of the system, complicating pump-probe signals at the earliest pump-probe time delays and requiring careful attention to distinguish from absorption effects like spectral hole burning.\cite{BritoCruz_Shank_1988,Fleugel_Antonetti_1987,Joffre_Koch_1988,SchmittRink_Haug_1988}
These potential applications and effects motivate a careful survey of the modification of OHG by the OSE.
% Ultrafast optical gating of OHG has been achieved through induced currents.\cite{Ruzicka_Zhao_2012}

In this Letter we apply the OSE to OHG by exploring IR pump, harmonic probe spectroscopy of WS\textsubscript{2}.
This OHG-probe spectroscopy is an example of the emerging methods which extend the capabilities of traditional pump-probe methods by using higher order interactions for the pump and/or probe.\cite{Morrow_Wright_2019, Morrow_Wright_2018, Xiong_Zanni_2009, Dietze_Mathies_2016, Mandal_Wasielewski_2019, Langer_Huber_2018}
If these methods are to be viable, it is imperative to understand how the OSE influences an OHG probe because important physics like charge separation at a two dimensional (2D) heterojunction occurs during pump-probe overlap.\cite{Hong_Wang_2014}
The OSE and OHG are pertinent for 2D transition metal dichalcogenides (TMDCs), where resonant optical transitions are generally strong\cite{Li_Heinz_2013, Malard_dePaula_2013, Wang_Zhao_2013, Wang_Urbaszek_2015} and optical damage thresholds are large.\cite{Paradisanos_Stratakis_2014}
Weak probes have previously shown that TMDCs have a strong OSE due to large exciton transition dipoles and intrinsic quantum confinement.\cite{Sie_Gedik_2014, Sie_Gedik_2016, Sie_Gedik_2017, Kim_Wang_2014, Yong_Wang_2018, Yong_Wang_2019, LaMountain_Stern_2018, Cunningham_Jonker_2019}
Herein we show that the OSE is strong for OHG, with resonance shift rates in excess of $2 \ \text{meV}$ per $\frac{\text{V}}{\text{nm}}$ of applied optical field.
Importantly, the precise sensitivity of the OSE and its competition with multiphoton absorption depends on morphology factors such as uniformity and thickness.
In addition to the well-known OSE blue shift, we find the OSE-OHG process incurs novel hybridization between the pump and probe fundamental fields.
When the pump and the probe fundamental have similar frequencies, quantum interference of the pump and probe photons strongly modulate the efficiency of OHG.
By tuning the pump frequency about the probe fundamental, the interference can either greatly suppress or enhance OHG.

Our experiments use two ultrafast optical parametric amplifiers (OPAs) to generate linearly polarized pump and probe electric fields ($\sim 50 \text{ fs}$ FWHM) (additional experimental details are available in the Supplemental Material (SM) \footnote{\SItxt}).
We measure the second harmonic generation (SHG) or third harmonic generation (THG) of the probe beam in the reflective direction (\autoref{fig:intro}c) from a single WS\textsubscript{2} screw-dislocation spiral (84 nm tall, \autoref{fig:intro}d,e) on a Si/SiO\textsubscript{2} substrate synthesized by water vapor assisted chemical vapor transport.\cite{Zhao_Jin_2019}
WS\textsubscript{2} on SiO\textsubscript{2} has a ground state A exciton transition at $E_0 \approx 1.98 \text{ eV}$. 
Though this work surveys many morphologies, a screw dislocation spiral of WS\textsubscript{2} is highlighted in the main text due to its bright SHG and THG.\cite{Shearer_Jin_2017, Fan_Pan_2017, Fan_Pan_2018}
\autoref{fig:intro}f shows the A exciton THG resonance of the spiral (blue line).
When a non-resonant pump (0.95 eV) is applied, the resonance blue-shifts (\autoref{fig:intro}f, green line), yielding an asymmetric difference lineshape (\autoref{fig:intro}g).

To investigate pump-OHG-probe, we measure the OHG dependence on the pump and probe frequency, relative arrival time, and fluence.
We look at changes in OHG intensity relative to the peak OHG of the unpumped spectrum:
\begin{equation}
	\text{norm. } \Delta I  \equiv \frac{I_{\text{OHG, pumped}} - I_{\text{OHG, unpumped}}}{ \text{max}\left\{I_{\text{OHG, unpumped}}\right\}}.\label{eq:normDi}
\end{equation}
The conventional pump-probe metric, in which signal is normalized by the probe \textit{spectrum} (i.e. $\Delta I / I$),\cite{Morrow_Wright_2019} is relinquished here because it over-emphasizes small changes at the wings of the OHG resonance, where the harmonic probe is weak.
Figures \ref{fig:intro}h,i show the pump-SHG(THG)-probe signal according to \autoref{eq:normDi} as the probe color and pump-probe delay are scanned.
The thick gray line traces the center of mass of the pumped OHG resonance at the different pump-probe time delays, $T$; the pump blue-shifts both SHG and THG with a time dependence that roughly follows the pump-probe temporal overlap.
The baseline value of the SHG gray line (1.92 eV) shows that its resonances is significantly shifted from that of the A exciton resonance (1.98 eV). 
This difference in resonance frequency, along with the significantly weaker transition dipole (see \cite{Note1}), suggests that the SHG resonance is not from the A exciton but perhaps an A trion.\cite{Seyler_Xu_2015}
This work will henceforth focus on the THG of the A exciton.

\begin{figure*}[!hbtp]
	\centering
	\includegraphics[trim={0cm 1cm 0cm 1.5cm},clip, width=\linewidth]{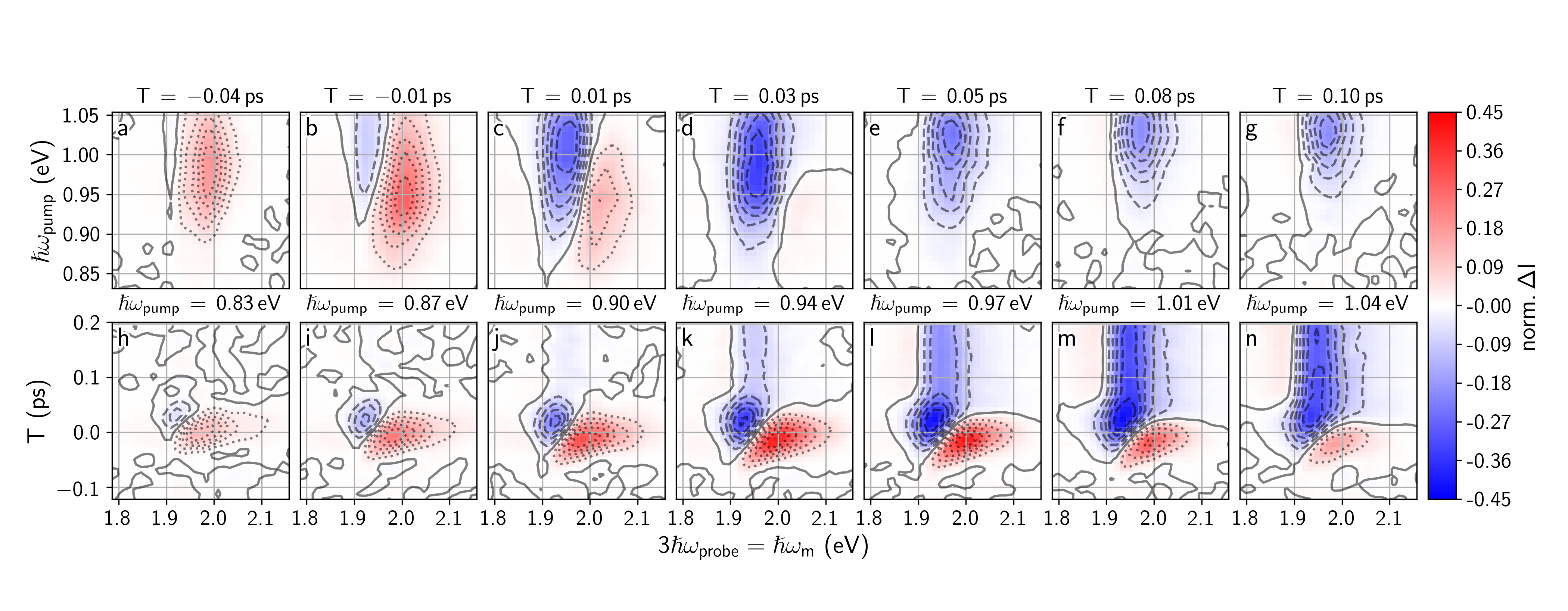}
	\caption{
		Effects of pump frequency, probe frequency, and pump-probe time delay on WS\textsubscript{2} THG spectrum.
		The first row shows pump frequency vs. probe frequency for seven time delays (noted in subfigure title).
		The second row shows time delay vs. probe frequency for seven pump frequencies (noted in subfigure title).
		The colormap is shared across all panels with contour lines locally normalized. $\mathcal{F}_{\text{pump}} \approx 3000 \frac{\mu \text{J}}{\text{cm}^2}$.}
	\label{fig:wigfreq} 
\end{figure*}

\autoref{fig:wigfreq} shows the THG dependence on pump frequency $\hbar\omega_{\text{pump}}$, probe frequency $\hbar\omega_{\text{probe}}$, and pump-probe time delay, $T$.
When the probe arrives before the pump ($T < 0$), THG is enhanced near the resonance ($\Delta I > 0$).
When pulses are overlapped ($T \approx 0$), the probe spectra (horizontal slices) are dispersive, which is consistent with blue-shifting of the exciton resonance.
When the probe is delayed by times greater than the pulse duration ($T > 50$ fs), response is observed only when $2\hbar\omega_{\text{pump}} > E_0$, indicating that the pump is dissipating energy via two photon absorption (2PA).
The effects of this absorption remain longer than the 10s of picoseconds experimentally accessible by our instrument (data shown in the SM\cite{Note1}).

Significantly, some of the pump-OHG-probe behaviors shown in \autoref{fig:wigfreq} run counter to expectations of the conventional OSE.
For example, the probe line shapes (horizontal slices of \autoref{fig:wigfreq}) are not strictly antisymmetric, contrary to expectations of a resonance shift.\cite{Knox_SchmittRink_1989, Yang_Beard_2016, Proppe_Sargent_2020}
The balance of the positive (red) and negative (blue) lobe is unequal and depends on the pump color, and the dominant lobe differs between SHG (stronger negative) and THG (stronger positive) probes under the same pump excitation (cf. Figures \ref{fig:intro}h and i).
Furthermore, the probe spectrum is strongly non-symmetric about $T=0$ (see \autoref{fig:wigfreq}h-n) which is counter to the expectation of adiabatic following of the pulse envelope.
These unusual behaviors cannot be explained by the incoherent population contributions because these contributions are negligible for many pump colors (\autoref{fig:wigfreq}h-j).

\begin{figure}[!htbp]
	\centering
	\includegraphics[trim={0cm 1cm 0cm .5cm},clip,width=\linewidth]{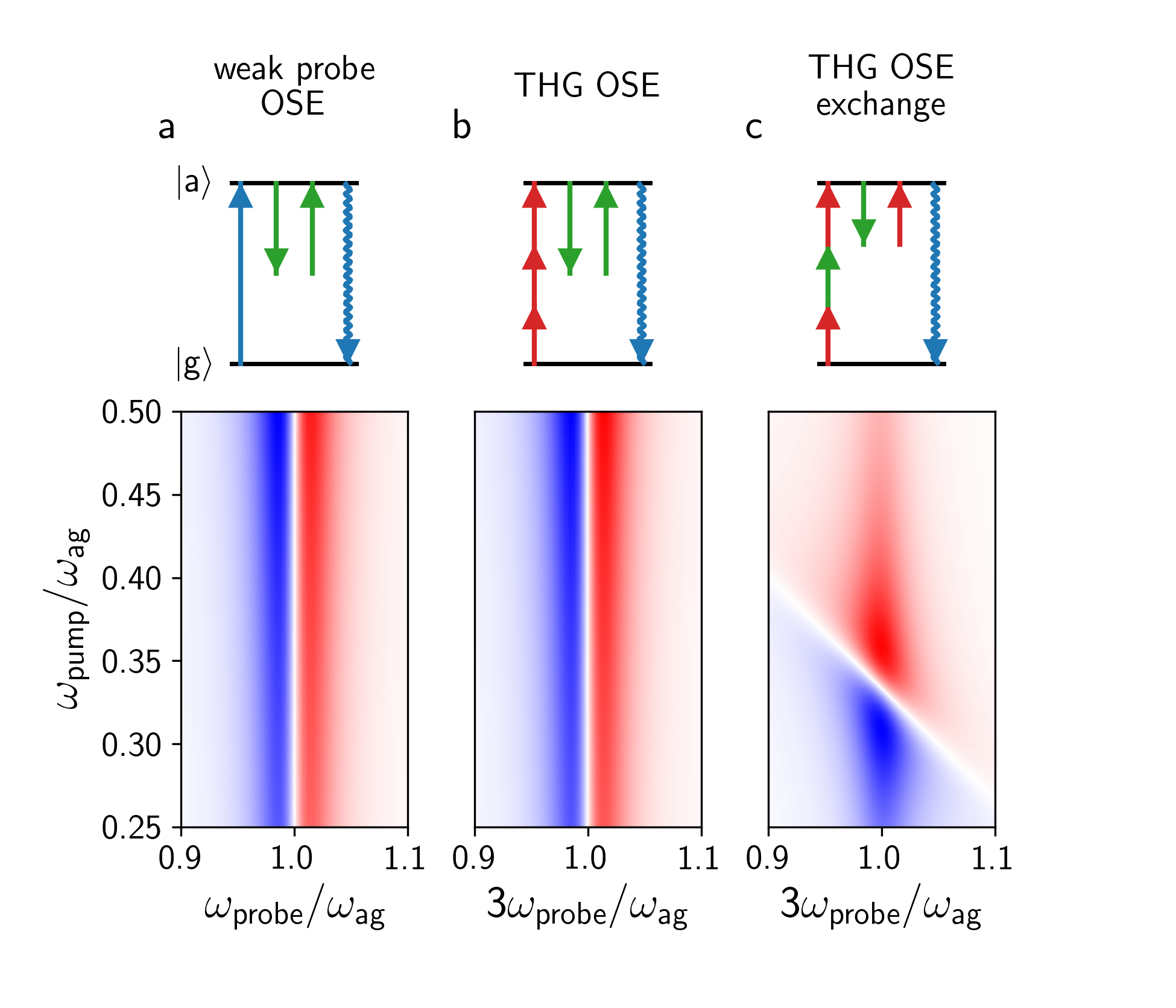}
	\caption{
		Representations of the OSE using wave-mixing energy level (WMEL\cite{Lee_Albrecht_1985}) diagrams (first row) and their corresponding 2D frequency response (second row).
		Straight arrows represent interactions of the input fields: the resonant weak field probe (blue),the detuned pump (green), and the probe fundamental (red).
		The wavy blue arrow represents the emission of the signal field.  
		Time flows from left to right.
		The 2D responses use $\Gamma = 0.025 \omega_{ag}$.
		The weak probe OSE response is calculated assuming a transient absorption measurement, while the THG responses assume \autoref{eq:normDi} (see the SM\cite{Note1} for details).
	}
	\label{fig:representative_WMELs} 
\end{figure}

To understand the differences between the conventional OSE and its manifestation in harmonic generation, we employed the well-known perturbative expansion technique to a two-level system.\cite{Mukamel_1999, Boyd_2008}
This expansion technique determines the non-linear response through a series of coherent, time-ordered, linear interactions with the pump and probe.
\autoref{fig:representative_WMELs}a shows this technique's diagrammatic representation of the OSE in a transient absorption measurement and its pump-probe frequency response.
The perturbative technique representation differs from \autoref{fig:intro}b in that it does not explicitly solve for the hybridized states; it does, however, explicitly treat the probe polarization.
As a consequence, the extension of the perturbative method to OHG is trivial, though cumbersome.
We provide a thorough walk-through of our treatment in the SM \cite{Note1}.
Although we focus on THG, the underlying effects are general to OHG.

The perturbative treatment recovers the OSE analogue for our experiment (THG OSE, \autoref{fig:representative_WMELs}b), in which the third harmonic polarization is generated and the pump subsequently drives the system.
An additional process arises, however, due to the similar frequencies of the pump and probe fundamental. 
When the probe fundamental and the pump have similar frequencies, they become indistinguishable, so their roles in harmonic generation and dressing the system can exchange.
A new pathway results (THG OSE exchange, \autoref{fig:representative_WMELs}c) where a triple-sum frequency (TSF,\cite{Morrow_Wright_2018} $2\omega_\text{probe} + \omega_\text{pump}$) polarization is dressed by both the probe and pump fields.
This pathway's 2D spectrum can be understood as a blue-shift of the TSF resonance, which explains the negatively sloped node.
This manifestation of the OSE is unique to harmonic generation because in the weak probe case, degeneracy of the pump and probe frequencies implies the pump is at resonance, where incoherent excitation (carrier populations) or strong field Rabi cycling effects will dominate.
Note that this approach is an example of quantum interference,\cite{Lin_Lupton_2019} which is related to recent work showing that photocurrents can be controled using interference of multiple multiphoton absorption processes.\cite{Wang_Cundiff_2019, Mahon_Sipe_2019, Muniz_Sipe_2019}

At $T=0$, the 2D spectrum for $n$-th harmonic generation of a two-level system can be approximated as:
\begin{equation}
\text{norm. } \Delta I  = \left| \frac{\Gamma \mu_{ag}\mathcal{E}_\text{pump}}{\Delta_{ag}^{(\text{OHG})}} \right|^2 \text{Re}\left[
\frac{x}{\Delta_{ag}^{(\text{OHG})}} + \frac{ny}{\Delta_{ag}^{(\text{exch})}}
\right], \label{eq:twostate_OSE_THG}
\end{equation} 
in which $\Gamma$ is the dephasing rate, resonance enhancement is determined by:
\begin{align}
\Delta_{ag}^{(\text{OHG})} &\equiv \hbar(\omega_{ag} - n\omega_\text{probe} - i\Gamma), \\
\Delta_{ag}^{(\text{exch})} &\equiv \hbar(\omega_{ag} - (n-1) \omega_\text{probe} - \omega_\text{pump} - i\Gamma),
\end{align}
and the quantities $x \equiv -\hbar^{-1} (n\omega_\text{probe} - \omega_\text{pump} -i/\Delta_t)^{-1}$ and $y \equiv - \hbar^{-1} ((n-1)\omega_\text{probe} + i/ \Delta_t)^{-1}$ are attenuation factors due to non-resonance for characteristic pulse durations of $\Delta_t$.
For a strongly detuned pump, $x$ and $y$ are relatively insensitive to pump and probe frequency, approximately equal in magnitude, and primarily real in character. 
The first term inside the real operator of \autoref{eq:twostate_OSE_THG} represents THG OSE,
while the second term represents THG OSE exchange.
The factor of $n$ enhancement of the exchange pathway is due to permutation symmetries of the pump and probe fields.

\begin{figure}[!htbp]
	\centering
	\includegraphics[trim={0cm 1cm 0cm 1.5cm},clip, width=\linewidth]{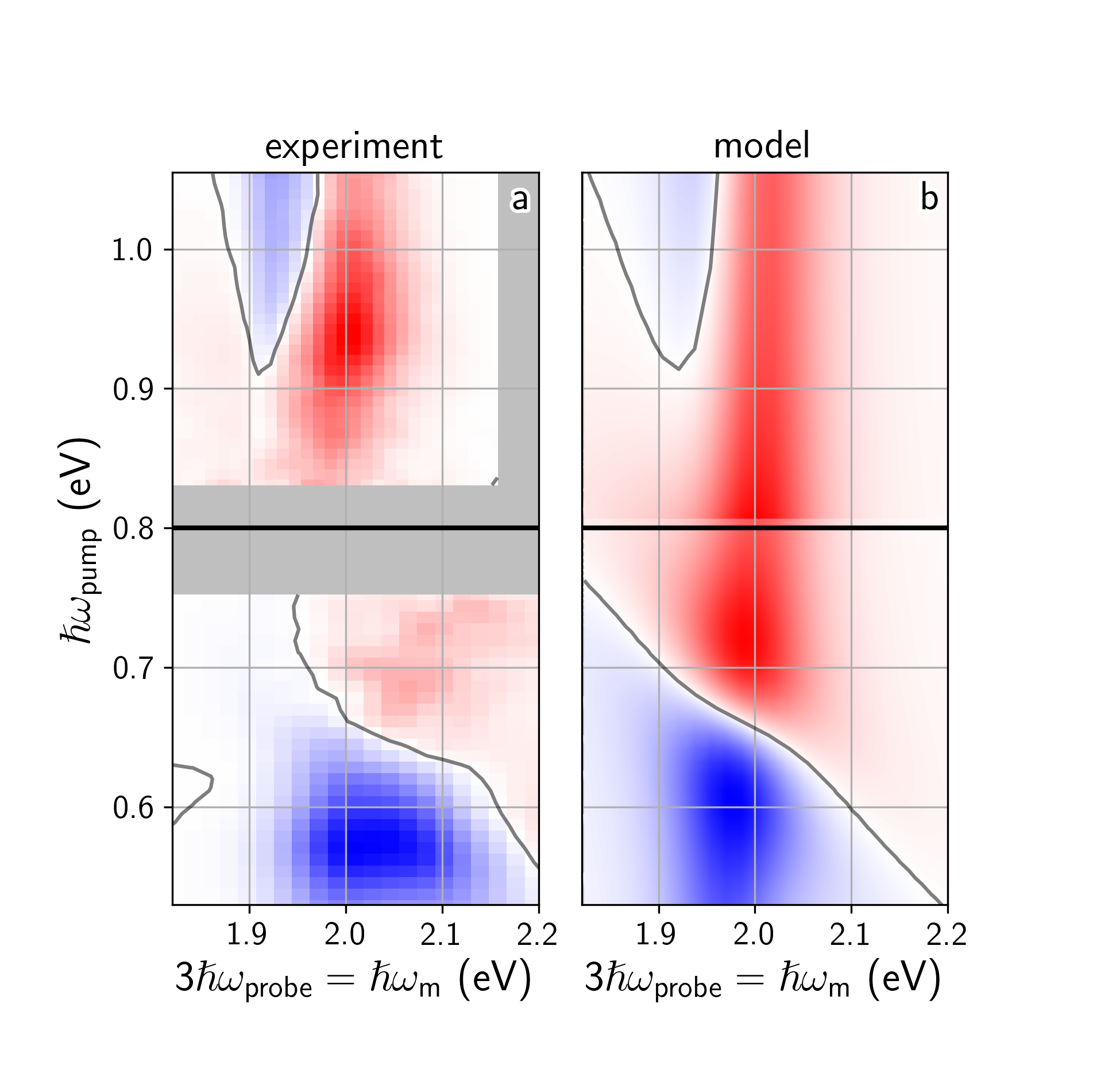}
	\caption{
		Comparison between experiment and perturbative expansion model of the OSE for THG including finite pulse effects ($\hbar\omega_{ag} = 1.98 \text{ eV}$, $\hbar\Gamma = 36 \text{ meV} \Rightarrow T_2=18 \text{ fs}$, and $\Delta_t=50\text{ fs}$ [the simulation is insensitive to the exact value of $\Delta_t$]).
		Experimental data are for $T=-0.01 \text{ ps}$. 
		The upper and lower halves of (a) and (b) each have their own colormap extent with red (blue) being an increase (decrease) in harmonic generation upon pump excitation.
		The gray section in (a) was not explored due to experimental constraints.
		Data in (a) were normalized along the pump axis by the frequency-dependent pump intensity.
	}
	\label{fig:sim_basic}
\end{figure}

\autoref{fig:sim_basic} plots a simulation of pump-THG-probe and compares it with experiment.
The simulation uses a numerical integration technique to account for a small pump-probe delay,\cite{Kohler_Wright_2017,  Gelin_Domcke_2005, WrightSim} which is qualitatively similar to \autoref{eq:twostate_OSE_THG}.
The simulation agrees well with experiment and the line shape can be understood as the weighted sum of the two THG OSE pathways in \autoref{fig:representative_WMELs}.
When pump and probe frequencies differ greatly (e.g. $\hbar\omega_\text{pump} \approx 1$ eV), the THG OSE effect is clearly resolved, and a blue shift along the probe axis is observed.
Near pump-probe degeneracy, however, the exchange pathway is prevalent, and a blue-shift normal to the $\omega_\text{pump} = \omega_{ag} - 2\omega_\text{probe}$ TSF resonance is seen.
Importantly, the blue shift from the exchange pathway enhances THG for pump frequencies greater than degeneracy and suppresses THG for pump frequencies less than degeneracy.
This observation explains the spectral asymmetry of the observed OHG OSE for both the cases of THG (e.g. \autoref{fig:wigfreq}) and SHG (e.g. \autoref{fig:intro}h).

Although this simulation shows the importance of the quantum interference between pathways, it only roughly reproduces the asymmetric dynamics about $T=0$ (compare \autoref{fig:wigfreq} to Figures S8 and S9 in the SM\cite{Note1}).
We believe this difference results because non-perturbative, higher-order effects must be taken into account to fully reproduce the dynamics in \autoref{fig:wigfreq}h-n.\cite{Blum_Kelley_1993}
Higher order effects are evidenced in our data by the pump induced 2PA and a probe-induced OSE (7 meV blueshift, Figure S11 in the SM\cite{Note1}), indicating an eight-wave mixing formalism (or higher) will be required to fully account for the asymmetric dynamics, which will be a goal for future work.

Dissipative coupling of the pump, through multi-photon absorption, competes with the coherent OSE and gives the long-lived ``bleach'' signals clearly seen at delays longer than pulse overlap, $T > 100$ fs. 
Multiphoton absorption is interesting for two opposing reasons: for pump-probe applications, it is a useful excitation mechanism because it reduces excitation pulse scatter;\cite{Morrow_Wright_2019} for ultrafast modulation applications, it diminishes the time resolution from the OSE, so it may be a parameter to minimize.
For optimizing ultrafast OSE modulation, it is beneficial to keep the pump color below the 2PA pump threshold $\hbar\omega_\text{pump} < E_0 / 2$ (cf. \autoref{fig:wigfreq}g) and close to pump and probe fundamental degeneracy, where the OSE exchange is enhanced (cf. \autoref{fig:sim_basic}).
Our two level model suggests contrast is also increased by using higher-order harmonics; as $n$ increases, the relative contribution of the OSE exchange process increases (\autoref{eq:twostate_OSE_THG}), and the degeneracy point $\omega_\text{pump} = \omega_\text{probe} = \omega_{ag} / n$ occurs at frequencies well below the 2PA onset.
This prediction holds when comparing SHG and THG; the SHG OSE is hard to isolate because the resonance enhancement overlaps with the 2PA onset (see Figure S10 in the SM\cite{Note1}).
Theoretically, the OHG OSE exchange process would also be strong for $n>3$ and should be sensitive to a wide variety of state symmetries through even vs. odd harmonic orders.\cite{Ye_Zhang_2014, Wang_Urbaszek_2015, Seyler_Xu_2015}
For high enough harmonics, however, our description will break down because the mechanism of harmonic generation becomes non-perturbative.\cite{Liu_Reis_2016,Ghimire_Reis_2019}

\begin{figure}[!htbp]
	\centering
	\includegraphics[trim={0cm 1cm 0cm 1cm},clip, width=\linewidth]{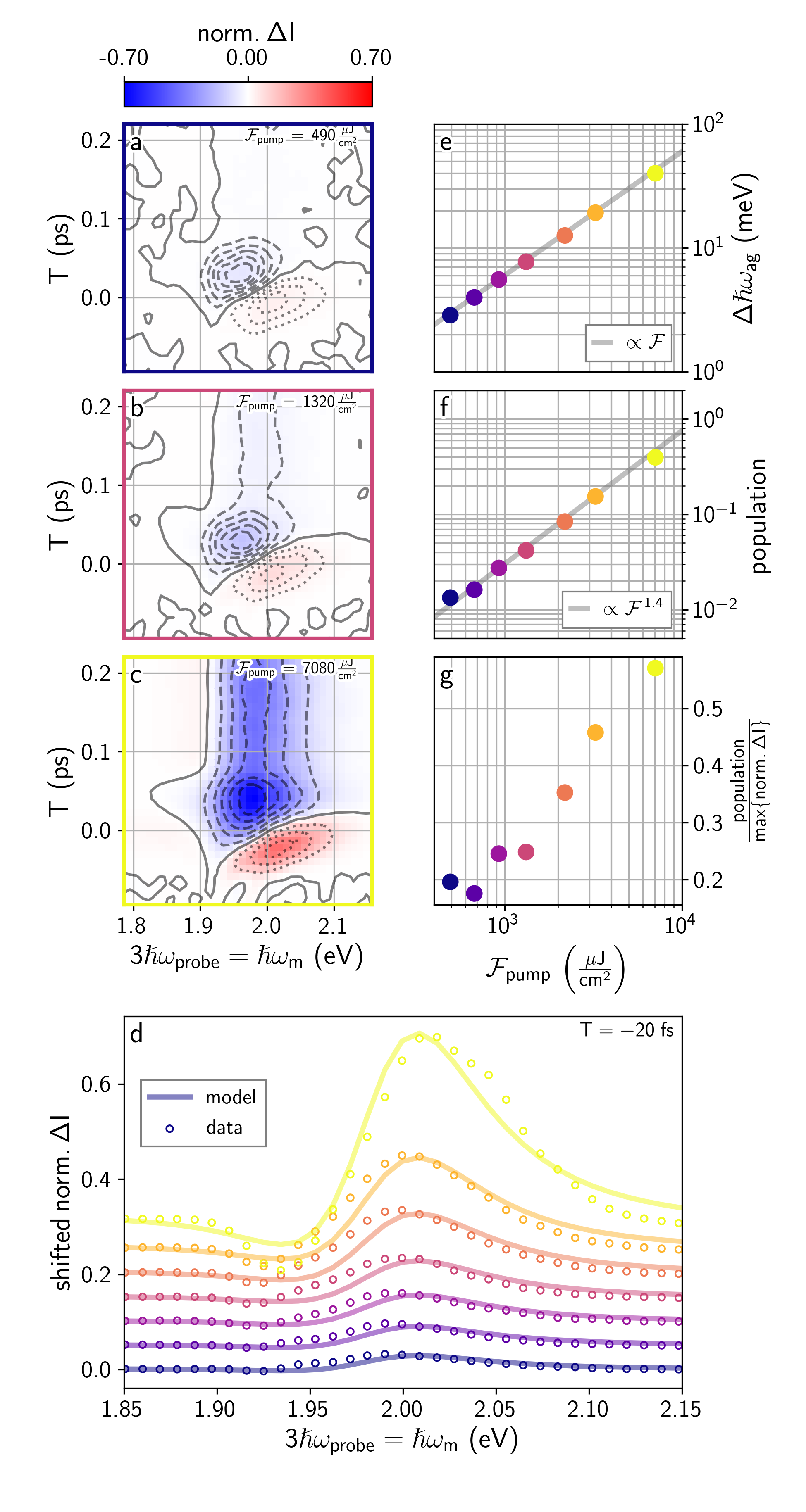}
	\caption{
		Effects of pump fluence and pump-probe time delay on WS\textsubscript{2} THG spectrum when $\hbar\omega_{\text{pump}}=0.99 \text{ eV}$.
		(a)-(c) Time delay vs. probe frequency for pump fluences of 490, 1320, and 7080 $\mu$J/cm\textsuperscript{2}.
		The colormap is shared across all panels, while contour lines are locally normalized.
		(d) $T=-20$ fs slices (color-keyed circles) at various pump fluences with lines showing the best fit of the data using \autoref{eq:twostate_OSE_THG}. 
		(e) $\Delta \hbar \omega_{ag}$ vs. pump fluence.  
		(f) Population response ($\text{peak norm. } \Delta I$  amplitude after 0.1 ps) vs. pump fluence showing a $\mathcal{F}^{1.4}$ trend.
		(g) The population response amplitude relative to the maximum amplitude near pulse overlap.
	}\label{fig:fluence}
\end{figure}

For pump-probe applications, it is interesting to further explore the nature of photoexcitation.
To this end, we performed fluence studies with the pump tuned to the 2PA threshold, $\hbar\omega_\text{pump} = 0.99$ eV.
\autoref{fig:fluence}a-c shows how THG changes for different pump fluences, $\mathcal{F}_{\text{pump}}$, ranging from 500 to 7000 $\frac{\mu \text{J}}{\text{cm}^2}$ (or, for our 50 fs pulse, electric field amplitudes ranging from roughly 1 to 20 $\frac{\text{V}}{\text{nm}}$).
At the lowest fluence used, signal is only seen near $T=0$; the coherent, short-lived OSE dominates (\autoref{fig:fluence}a).
As expected for a multi-photon mechanism, however, the prominence of the population signal increases with increased pump fluence.
At a sufficiently high fluence to suppress the majority ($\sim 70\%$) of the THG resonance the persistent population signal is almost as large as the peak signals near $T=0$ (\autoref{fig:fluence}c).

\autoref{fig:fluence}e-g plot the fluence scaling trends of the population and OSE signals.
To quantify the OSE, we fit the line shapes near $T=0$ to \autoref{eq:twostate_OSE_THG} to recover the Stark shift parameter, $2\hbar \Delta\omega_{ag} = |\mu \mathcal{E}_\text{pump}|^2 / (\omega_{ag} - \omega_\text{pump})$ (see the SM\cite{Note1} for details).
The fit quality is good (\autoref{fig:fluence}d), the shift is linear with fluence (\autoref{fig:fluence}e), and a maximum shift of $\sim 40$ meV is achieved.
To quantify the population signal, we took the maximum signal amplitude for $T > 100$ fs.  
The population response (\autoref{fig:fluence}g) scales as $\mathcal{F}_{\text{pump}}^{1.4}$.
Two-photon absorption is expected to scale as $\mathcal{F}_{\text{pump}}^{2}$, or to saturate at high fluence; the discrepancy in scaling behavior is not understood at this time, although the OSE is known to introduce surprising fluence scaling due to dynamic resonance conditions and OHG broadening.\cite{Elgin_New_1976}
It is also possible that these excitation densities achieved are large enough to introduce strong screening effects, which would also alter simple fluence scaling trends.
Finally, the contrast between the population and the OSE signals are measured by comparing the ratio of the peak amplitude of signals near zero delay with the peak amplitude at delays beyond 100 fs (\autoref{fig:fluence}g).

% DDK: we should consider revising the fluence figure to include data for how other morphologies behave
% DDK: my suggestion is to wait for Song to read this to get a better idea of what needs to be communicated.
We also measured multidimensional pump-THG-probe in a variety of WS$_2$ morphologies to further investigate the balance of absorptive and coherent processes (see Figures S11-S19 in the SM\cite{Note1}).
In all morphologies examined, the pump spectrum at $T>100$ fs revealed a 2PA onset.
Highly crystalline samples, such as a single monolayer, exhibit a sharp 2PA pump color onset at $\hbar\omega_\text{pump} = E_0 / 2$, a narrow THG resonance, and roughly quadratic fluence scaling of the population. 
% DDK: we could use another figure/subfigure that shows the pump frequency dependence at long times.
Less crystalline samples, such as a polycrystalline thin film, exhibit a broad 2PA pump color onset, a broad THG resonance, and roughly linear fluence scaling of the population.
The other samples examined were in between these behaviors.
The differences in absorption behaviors showcase the variety of behaviors a sub-band edge pump can induce.

In this work we investigated both coherent and incoherent alteration of resonant harmonic generation in WS\textsubscript{2} from an intense, sub-band edge pump.
The pump field not only shifts the resonant transition in a manner similar to the traditional optical Stark effect with a single photon probe, but also exchanges with the probe field, resulting in a novel way to shift and modulate the intensity of the harmonic output.
A simple perturbation theory approach adequately captures the measured multidimensional spectral and dynamic characteristics and will be essential for interpreting pump-OHG-probe spectra near the temporal overlap regime.
The OHG OSE exchange mechanisms may also have important applications in ultrafast photonic signal modulation. 

\vspace{\baselineskip}
All data and scripts used in this work are permissively licensed and available for reuse at \href{{http://dx.doi.org/10.17605/OSF.IO/SNTPC}}{DOI: 10.17605/OSF.IO/SNTPC}
We acknowledge support from the Department of Energy, Office of Basic Energy Sciences, Division of Materials Sciences and Engineering, under award DE-FG02-09ER46664.	
D.J.M. acknowledges support from the Link Foundation.

\bibliography{database}

\begin{thebibliography}{62}%
\makeatletter
\providecommand \@ifxundefined [1]{%
 \@ifx{#1\undefined}
}%
\providecommand \@ifnum [1]{%
 \ifnum #1\expandafter \@firstoftwo
 \else \expandafter \@secondoftwo
 \fi
}%
\providecommand \@ifx [1]{%
 \ifx #1\expandafter \@firstoftwo
 \else \expandafter \@secondoftwo
 \fi
}%
\providecommand \natexlab [1]{#1}%
\providecommand \enquote  [1]{``#1''}%
\providecommand \bibnamefont  [1]{#1}%
\providecommand \bibfnamefont [1]{#1}%
\providecommand \citenamefont [1]{#1}%
\providecommand \href@noop [0]{\@secondoftwo}%
\providecommand \href [0]{\begingroup \@sanitize@url \@href}%
\providecommand \@href[1]{\@@startlink{#1}\@@href}%
\providecommand \@@href[1]{\endgroup#1\@@endlink}%
\providecommand \@sanitize@url [0]{\catcode `\\12\catcode `\$12\catcode
  `\&12\catcode `\#12\catcode `\^12\catcode `\_12\catcode `\%12\relax}%
\providecommand \@@startlink[1]{}%
\providecommand \@@endlink[0]{}%
\providecommand \url  [0]{\begingroup\@sanitize@url \@url }%
\providecommand \@url [1]{\endgroup\@href {#1}{\urlprefix }}%
\providecommand \urlprefix  [0]{URL }%
\providecommand \Eprint [0]{\href }%
\providecommand \doibase [0]{https://doi.org/}%
\providecommand \selectlanguage [0]{\@gobble}%
\providecommand \bibinfo  [0]{\@secondoftwo}%
\providecommand \bibfield  [0]{\@secondoftwo}%
\providecommand \translation [1]{[#1]}%
\providecommand \BibitemOpen [0]{}%
\providecommand \bibitemStop [0]{}%
\providecommand \bibitemNoStop [0]{.\EOS\space}%
\providecommand \EOS [0]{\spacefactor3000\relax}%
\providecommand \BibitemShut  [1]{\csname bibitem#1\endcsname}%
\let\auto@bib@innerbib\@empty
%</preamble>
\bibitem [{\citenamefont {Franken}\ \emph {et~al.}(1961)\citenamefont
  {Franken}, \citenamefont {Hill}, \citenamefont {Peters},\ and\ \citenamefont
  {Weinreich}}]{Franken_Weinreich_1961}%
  \BibitemOpen
  \bibfield  {author} {\bibinfo {author} {\bibfnamefont {P.~A.}\ \bibnamefont
  {Franken}}, \bibinfo {author} {\bibfnamefont {A.~E.}\ \bibnamefont {Hill}},
  \bibinfo {author} {\bibfnamefont {C.~W.}\ \bibnamefont {Peters}},\ and\
  \bibinfo {author} {\bibfnamefont {G.}~\bibnamefont {Weinreich}},\ }\bibfield
  {title} {\bibinfo {title} {Generation of optical harmonics},\ }\href
  {https://doi.org/10.1103/physrevlett.7.118} {\bibfield  {journal} {\bibinfo
  {journal} {Phys. Rev. Lett.}\ }\textbf {\bibinfo {volume} {7}},\ \bibinfo
  {pages} {118} (\bibinfo {year} {1961})}\BibitemShut {NoStop}%
\bibitem [{\citenamefont {Maker}\ \emph {et~al.}(1962)\citenamefont {Maker},
  \citenamefont {Terhune}, \citenamefont {Nisenoff},\ and\ \citenamefont
  {Savage}}]{Maker_Savage_1962}%
  \BibitemOpen
  \bibfield  {author} {\bibinfo {author} {\bibfnamefont {P.~D.}\ \bibnamefont
  {Maker}}, \bibinfo {author} {\bibfnamefont {R.~W.}\ \bibnamefont {Terhune}},
  \bibinfo {author} {\bibfnamefont {M.}~\bibnamefont {Nisenoff}},\ and\
  \bibinfo {author} {\bibfnamefont {C.~M.}\ \bibnamefont {Savage}},\ }\bibfield
   {title} {\bibinfo {title} {Effects of dispersion and focusing on the
  production of optical harmonics},\ }\href
  {https://doi.org/10.1103/physrevlett.8.21} {\bibfield  {journal} {\bibinfo
  {journal} {Phys. Rev. Lett.}\ }\textbf {\bibinfo {volume} {8}},\ \bibinfo
  {pages} {21} (\bibinfo {year} {1962})}\BibitemShut {NoStop}%
\bibitem [{\citenamefont {Li}\ \emph {et~al.}(2013)\citenamefont {Li},
  \citenamefont {Rao}, \citenamefont {Mak}, \citenamefont {You}, \citenamefont
  {Wang}, \citenamefont {Dean},\ and\ \citenamefont {Heinz}}]{Li_Heinz_2013}%
  \BibitemOpen
  \bibfield  {author} {\bibinfo {author} {\bibfnamefont {Y.}~\bibnamefont
  {Li}}, \bibinfo {author} {\bibfnamefont {Y.}~\bibnamefont {Rao}}, \bibinfo
  {author} {\bibfnamefont {K.~F.}\ \bibnamefont {Mak}}, \bibinfo {author}
  {\bibfnamefont {Y.}~\bibnamefont {You}}, \bibinfo {author} {\bibfnamefont
  {S.}~\bibnamefont {Wang}}, \bibinfo {author} {\bibfnamefont {C.~R.}\
  \bibnamefont {Dean}},\ and\ \bibinfo {author} {\bibfnamefont {T.~F.}\
  \bibnamefont {Heinz}},\ }\bibfield  {title} {\bibinfo {title} {Probing
  symmetry properties of few-layer {MoS}$_2$ and h-{BN} by optical
  second-harmonic generation},\ }\href {https://doi.org/10.1021/nl401561r}
  {\bibfield  {journal} {\bibinfo  {journal} {Nano Lett.}\ }\textbf {\bibinfo
  {volume} {13}},\ \bibinfo {pages} {3329} (\bibinfo {year}
  {2013})}\BibitemShut {NoStop}%
\bibitem [{\citenamefont {Seyler}\ \emph {et~al.}(2015)\citenamefont {Seyler},
  \citenamefont {Schaibley}, \citenamefont {Gong}, \citenamefont {Rivera},
  \citenamefont {Jones}, \citenamefont {Wu}, \citenamefont {Yan}, \citenamefont
  {Mandrus}, \citenamefont {Yao},\ and\ \citenamefont {Xu}}]{Seyler_Xu_2015}%
  \BibitemOpen
  \bibfield  {author} {\bibinfo {author} {\bibfnamefont {K.~L.}\ \bibnamefont
  {Seyler}}, \bibinfo {author} {\bibfnamefont {J.~R.}\ \bibnamefont
  {Schaibley}}, \bibinfo {author} {\bibfnamefont {P.}~\bibnamefont {Gong}},
  \bibinfo {author} {\bibfnamefont {P.}~\bibnamefont {Rivera}}, \bibinfo
  {author} {\bibfnamefont {A.~M.}\ \bibnamefont {Jones}}, \bibinfo {author}
  {\bibfnamefont {S.}~\bibnamefont {Wu}}, \bibinfo {author} {\bibfnamefont
  {J.}~\bibnamefont {Yan}}, \bibinfo {author} {\bibfnamefont {D.~G.}\
  \bibnamefont {Mandrus}}, \bibinfo {author} {\bibfnamefont {W.}~\bibnamefont
  {Yao}},\ and\ \bibinfo {author} {\bibfnamefont {X.}~\bibnamefont {Xu}},\
  }\bibfield  {title} {\bibinfo {title} {Electrical control of second-harmonic
  generation in a {WSe}$_2$ monolayer transistor},\ }\href
  {https://doi.org/10.1038/nnano.2015.73} {\bibfield  {journal} {\bibinfo
  {journal} {Nat. Nanotechnol.}\ }\textbf {\bibinfo {volume} {10}},\ \bibinfo
  {pages} {407} (\bibinfo {year} {2015})}\BibitemShut {NoStop}%
\bibitem [{\citenamefont {Sirtori}\ \emph {et~al.}(1992)\citenamefont
  {Sirtori}, \citenamefont {Capasso}, \citenamefont {Sivco}, \citenamefont
  {Hutchinson},\ and\ \citenamefont {Cho}}]{Sirtori_Cho_1992}%
  \BibitemOpen
  \bibfield  {author} {\bibinfo {author} {\bibfnamefont {C.}~\bibnamefont
  {Sirtori}}, \bibinfo {author} {\bibfnamefont {F.}~\bibnamefont {Capasso}},
  \bibinfo {author} {\bibfnamefont {D.~L.}\ \bibnamefont {Sivco}}, \bibinfo
  {author} {\bibfnamefont {A.~L.}\ \bibnamefont {Hutchinson}},\ and\ \bibinfo
  {author} {\bibfnamefont {A.~Y.}\ \bibnamefont {Cho}},\ }\bibfield  {title}
  {\bibinfo {title} {Resonant stark tuning of second-order susceptibility in
  coupled quantum wells},\ }\href {https://doi.org/10.1063/1.106999} {\bibfield
   {journal} {\bibinfo  {journal} {Appl. Phys. Lett.}\ }\textbf {\bibinfo
  {volume} {60}},\ \bibinfo {pages} {151} (\bibinfo {year} {1992})}\BibitemShut
  {NoStop}%
\bibitem [{\citenamefont {Lafrentz}\ \emph {et~al.}(2013)\citenamefont
  {Lafrentz}, \citenamefont {Brunne}, \citenamefont {Kaminski}, \citenamefont
  {Pavlov}, \citenamefont {Rodina}, \citenamefont {Pisarev}, \citenamefont
  {Yakovlev}, \citenamefont {Bakin},\ and\ \citenamefont
  {Bayer}}]{Lafrentz_Bayer_2013}%
  \BibitemOpen
  \bibfield  {author} {\bibinfo {author} {\bibfnamefont {M.}~\bibnamefont
  {Lafrentz}}, \bibinfo {author} {\bibfnamefont {D.}~\bibnamefont {Brunne}},
  \bibinfo {author} {\bibfnamefont {B.}~\bibnamefont {Kaminski}}, \bibinfo
  {author} {\bibfnamefont {V.~V.}\ \bibnamefont {Pavlov}}, \bibinfo {author}
  {\bibfnamefont {A.~V.}\ \bibnamefont {Rodina}}, \bibinfo {author}
  {\bibfnamefont {R.~V.}\ \bibnamefont {Pisarev}}, \bibinfo {author}
  {\bibfnamefont {D.~R.}\ \bibnamefont {Yakovlev}}, \bibinfo {author}
  {\bibfnamefont {A.}~\bibnamefont {Bakin}},\ and\ \bibinfo {author}
  {\bibfnamefont {M.}~\bibnamefont {Bayer}},\ }\bibfield  {title} {\bibinfo
  {title} {Magneto-stark effect of excitons as the origin of second harmonic
  generation in {ZnO}},\ }\href
  {https://doi.org/10.1103/physrevlett.110.116402} {\bibfield  {journal}
  {\bibinfo  {journal} {Phys. Rev. Lett.}\ }\textbf {\bibinfo {volume} {110}},\
  \bibinfo {pages} {116402} (\bibinfo {year} {2013})}\BibitemShut {NoStop}%
\bibitem [{\citenamefont {Brunne}\ \emph {et~al.}(2015)\citenamefont {Brunne},
  \citenamefont {Lafrentz}, \citenamefont {Pavlov}, \citenamefont {Pisarev},
  \citenamefont {Rodina}, \citenamefont {Yakovlev},\ and\ \citenamefont
  {Bayer}}]{Brunne_Bayer_2015}%
  \BibitemOpen
  \bibfield  {author} {\bibinfo {author} {\bibfnamefont {D.}~\bibnamefont
  {Brunne}}, \bibinfo {author} {\bibfnamefont {M.}~\bibnamefont {Lafrentz}},
  \bibinfo {author} {\bibfnamefont {V.~V.}\ \bibnamefont {Pavlov}}, \bibinfo
  {author} {\bibfnamefont {R.~V.}\ \bibnamefont {Pisarev}}, \bibinfo {author}
  {\bibfnamefont {A.~V.}\ \bibnamefont {Rodina}}, \bibinfo {author}
  {\bibfnamefont {D.~R.}\ \bibnamefont {Yakovlev}},\ and\ \bibinfo {author}
  {\bibfnamefont {M.}~\bibnamefont {Bayer}},\ }\bibfield  {title} {\bibinfo
  {title} {Electric field effect on optical harmonic generation at the exciton
  resonances in {GaAs}},\ }\href {https://doi.org/10.1103/physrevb.92.085202}
  {\bibfield  {journal} {\bibinfo  {journal} {Phys. Rev. B}\ }\textbf {\bibinfo
  {volume} {92}},\ \bibinfo {pages} {085202} (\bibinfo {year}
  {2015})}\BibitemShut {NoStop}%
\bibitem [{\citenamefont {Wang}\ \emph {et~al.}(2015)\citenamefont {Wang},
  \citenamefont {Marie}, \citenamefont {Gerber}, \citenamefont {Amand},
  \citenamefont {Lagarde}, \citenamefont {Bouet}, \citenamefont {Vidal},
  \citenamefont {Balocchi},\ and\ \citenamefont
  {Urbaszek}}]{Wang_Urbaszek_2015}%
  \BibitemOpen
  \bibfield  {author} {\bibinfo {author} {\bibfnamefont {G.}~\bibnamefont
  {Wang}}, \bibinfo {author} {\bibfnamefont {X.}~\bibnamefont {Marie}},
  \bibinfo {author} {\bibfnamefont {I.}~\bibnamefont {Gerber}}, \bibinfo
  {author} {\bibfnamefont {T.}~\bibnamefont {Amand}}, \bibinfo {author}
  {\bibfnamefont {D.}~\bibnamefont {Lagarde}}, \bibinfo {author} {\bibfnamefont
  {L.}~\bibnamefont {Bouet}}, \bibinfo {author} {\bibfnamefont
  {M.}~\bibnamefont {Vidal}}, \bibinfo {author} {\bibfnamefont
  {A.}~\bibnamefont {Balocchi}},\ and\ \bibinfo {author} {\bibfnamefont
  {B.}~\bibnamefont {Urbaszek}},\ }\bibfield  {title} {\bibinfo {title} {Giant
  enhancement of the optical second-harmonic emission of {WSe}$_2$ monolayers
  by laser excitation at exciton resonances},\ }\href
  {https://doi.org/10.1103/physrevlett.114.097403} {\bibfield  {journal}
  {\bibinfo  {journal} {Phys. Rev. Lett.}\ }\textbf {\bibinfo {volume} {114}},\
  \bibinfo {pages} {097403} (\bibinfo {year} {2015})}\BibitemShut {NoStop}%
\bibitem [{\citenamefont {Wang}\ and\ \citenamefont
  {Davis}(1975)}]{Wang_Davis_1975}%
  \BibitemOpen
  \bibfield  {author} {\bibinfo {author} {\bibfnamefont {C.~C.}\ \bibnamefont
  {Wang}}\ and\ \bibinfo {author} {\bibfnamefont {L.~I.}\ \bibnamefont
  {Davis}},\ }\bibfield  {title} {\bibinfo {title} {Saturation of resonant
  two-photon transitions in thallium vapor},\ }\href
  {https://doi.org/10.1103/PhysRevLett.35.650} {\bibfield  {journal} {\bibinfo
  {journal} {Phys. Rev. Lett.}\ }\textbf {\bibinfo {volume} {35}},\ \bibinfo
  {pages} {650} (\bibinfo {year} {1975})}\BibitemShut {NoStop}%
\bibitem [{\citenamefont {Morrow}\ \emph {et~al.}(2018)\citenamefont {Morrow},
  \citenamefont {Kohler}, \citenamefont {Czech},\ and\ \citenamefont
  {Wright}}]{Morrow_Wright_2018}%
  \BibitemOpen
  \bibfield  {author} {\bibinfo {author} {\bibfnamefont {D.~J.}\ \bibnamefont
  {Morrow}}, \bibinfo {author} {\bibfnamefont {D.~D.}\ \bibnamefont {Kohler}},
  \bibinfo {author} {\bibfnamefont {K.~J.}\ \bibnamefont {Czech}},\ and\
  \bibinfo {author} {\bibfnamefont {J.~C.}\ \bibnamefont {Wright}},\ }\bibfield
   {title} {\bibinfo {title} {Communication: Multidimensional triple
  sum-frequency spectroscopy of {MoS}$_2$ and comparisons with absorption and
  second harmonic generation spectroscopies},\ }\href
  {https://doi.org/10.1063/1.5047802} {\bibfield  {journal} {\bibinfo
  {journal} {J. Chem. Phys.}\ }\textbf {\bibinfo {volume} {149}},\ \bibinfo
  {pages} {091101} (\bibinfo {year} {2018})}\BibitemShut {NoStop}%
\bibitem [{\citenamefont {Morrow}\ \emph {et~al.}(2019)\citenamefont {Morrow},
  \citenamefont {Kohler}, \citenamefont {Zhao}, \citenamefont {Jin},\ and\
  \citenamefont {Wright}}]{Morrow_Wright_2019}%
  \BibitemOpen
  \bibfield  {author} {\bibinfo {author} {\bibfnamefont {D.~J.}\ \bibnamefont
  {Morrow}}, \bibinfo {author} {\bibfnamefont {D.~D.}\ \bibnamefont {Kohler}},
  \bibinfo {author} {\bibfnamefont {Y.}~\bibnamefont {Zhao}}, \bibinfo {author}
  {\bibfnamefont {S.}~\bibnamefont {Jin}},\ and\ \bibinfo {author}
  {\bibfnamefont {J.~C.}\ \bibnamefont {Wright}},\ }\bibfield  {title}
  {\bibinfo {title} {Triple sum frequency pump-probe spectroscopy of transition
  metal dichalcogenides},\ }\href {https://doi.org/10.1103/physrevb.100.235303}
  {\bibfield  {journal} {\bibinfo  {journal} {Phys. Rev. B}\ }\textbf {\bibinfo
  {volume} {100}},\ \bibinfo {pages} {235303} (\bibinfo {year}
  {2019})}\BibitemShut {NoStop}%
\bibitem [{\citenamefont {Autler}\ and\ \citenamefont
  {Townes}(1955)}]{Autler_Townes_1955}%
  \BibitemOpen
  \bibfield  {author} {\bibinfo {author} {\bibfnamefont {S.~H.}\ \bibnamefont
  {Autler}}\ and\ \bibinfo {author} {\bibfnamefont {C.~H.}\ \bibnamefont
  {Townes}},\ }\bibfield  {title} {\bibinfo {title} {Stark effect in rapidly
  varying fields},\ }\href {https://doi.org/10.1103/physrev.100.703} {\bibfield
   {journal} {\bibinfo  {journal} {Phys. Rev.}\ }\textbf {\bibinfo {volume}
  {100}},\ \bibinfo {pages} {703} (\bibinfo {year} {1955})}\BibitemShut
  {NoStop}%
\bibitem [{\citenamefont {Bakos}(1977)}]{Bakos_1977}%
  \BibitemOpen
  \bibfield  {author} {\bibinfo {author} {\bibfnamefont {J.}~\bibnamefont
  {Bakos}},\ }\bibfield  {title} {\bibinfo {title} {{AC} stark effect and
  multiphoton processes in atoms},\ }\href
  {https://doi.org/10.1016/0370-1573(77)90016-3} {\bibfield  {journal}
  {\bibinfo  {journal} {Phys. Rep.}\ }\textbf {\bibinfo {volume} {31}},\
  \bibinfo {pages} {209} (\bibinfo {year} {1977})}\BibitemShut {NoStop}%
\bibitem [{\citenamefont {Sussman}(2011)}]{Sussman_2011}%
  \BibitemOpen
  \bibfield  {author} {\bibinfo {author} {\bibfnamefont {B.~J.}\ \bibnamefont
  {Sussman}},\ }\bibfield  {title} {\bibinfo {title} {Five ways to the
  nonresonant dynamic stark effect},\ }\href
  {https://doi.org/10.1119/1.3553018} {\bibfield  {journal} {\bibinfo
  {journal} {Am. J. Phys.}\ }\textbf {\bibinfo {volume} {79}},\ \bibinfo
  {pages} {477} (\bibinfo {year} {2011})}\BibitemShut {NoStop}%
\bibitem [{\citenamefont {Boyd}(2008)}]{Boyd_2008}%
  \BibitemOpen
  \bibfield  {author} {\bibinfo {author} {\bibfnamefont {R.~W.}\ \bibnamefont
  {Boyd}},\ }\href@noop {} {\emph {\bibinfo {title} {Nonlinear Optics}}},\
  \bibinfo {edition} {3rd}\ ed.\ (\bibinfo  {publisher} {Academic Press},\
  \bibinfo {year} {2008})\BibitemShut {NoStop}%
\bibitem [{\citenamefont {Mysyrowicz}\ \emph {et~al.}(1986)\citenamefont
  {Mysyrowicz}, \citenamefont {Hulin}, \citenamefont {Antonetti}, \citenamefont
  {Migus}, \citenamefont {Masselink},\ and\ \citenamefont
  {Morko\ifmmode~\mbox{\c{c}}\else \c{c}\fi{}}}]{Mysyrowicz_1986}%
  \BibitemOpen
  \bibfield  {author} {\bibinfo {author} {\bibfnamefont {A.}~\bibnamefont
  {Mysyrowicz}}, \bibinfo {author} {\bibfnamefont {D.}~\bibnamefont {Hulin}},
  \bibinfo {author} {\bibfnamefont {A.}~\bibnamefont {Antonetti}}, \bibinfo
  {author} {\bibfnamefont {A.}~\bibnamefont {Migus}}, \bibinfo {author}
  {\bibfnamefont {W.~T.}\ \bibnamefont {Masselink}},\ and\ \bibinfo {author}
  {\bibfnamefont {H.}~\bibnamefont {Morko\ifmmode~\mbox{\c{c}}\else
  \c{c}\fi{}}},\ }\bibfield  {title} {\bibinfo {title} {"dressed excitons" in a
  multiple-quantum-well structure: Evidence for an optical stark effect with
  femtosecond response time},\ }\href
  {https://doi.org/10.1103/PhysRevLett.56.2748} {\bibfield  {journal} {\bibinfo
   {journal} {Phys. Rev. Lett.}\ }\textbf {\bibinfo {volume} {56}},\ \bibinfo
  {pages} {2748} (\bibinfo {year} {1986})}\BibitemShut {NoStop}%
\bibitem [{\citenamefont {Peyghambarian}\ \emph {et~al.}(1989)\citenamefont
  {Peyghambarian}, \citenamefont {Koch}, \citenamefont {Lindberg},
  \citenamefont {Fluegel},\ and\ \citenamefont
  {Joffre}}]{Peyghambarian_Joffre_1989}%
  \BibitemOpen
  \bibfield  {author} {\bibinfo {author} {\bibfnamefont {N.}~\bibnamefont
  {Peyghambarian}}, \bibinfo {author} {\bibfnamefont {S.~W.}\ \bibnamefont
  {Koch}}, \bibinfo {author} {\bibfnamefont {M.}~\bibnamefont {Lindberg}},
  \bibinfo {author} {\bibfnamefont {B.}~\bibnamefont {Fluegel}},\ and\ \bibinfo
  {author} {\bibfnamefont {M.}~\bibnamefont {Joffre}},\ }\bibfield  {title}
  {\bibinfo {title} {Dynamic stark effect of exciton and continuum states in
  {CdS}},\ }\href {https://doi.org/10.1103/physrevlett.62.1185} {\bibfield
  {journal} {\bibinfo  {journal} {Phys. Rev. Lett.}\ }\textbf {\bibinfo
  {volume} {62}},\ \bibinfo {pages} {1185} (\bibinfo {year}
  {1989})}\BibitemShut {NoStop}%
\bibitem [{\citenamefont {Knox}\ \emph {et~al.}(1989)\citenamefont {Knox},
  \citenamefont {Chemla}, \citenamefont {Miller}, \citenamefont {Stark},\ and\
  \citenamefont {Schmitt-Rink}}]{Knox_SchmittRink_1989}%
  \BibitemOpen
  \bibfield  {author} {\bibinfo {author} {\bibfnamefont {W.~H.}\ \bibnamefont
  {Knox}}, \bibinfo {author} {\bibfnamefont {D.~S.}\ \bibnamefont {Chemla}},
  \bibinfo {author} {\bibfnamefont {D.~A.~B.}\ \bibnamefont {Miller}}, \bibinfo
  {author} {\bibfnamefont {J.~B.}\ \bibnamefont {Stark}},\ and\ \bibinfo
  {author} {\bibfnamefont {S.}~\bibnamefont {Schmitt-Rink}},\ }\bibfield
  {title} {\bibinfo {title} {Femtosecond ac stark effect in semiconductor
  quantum wells: Extreme low- and high-intensity limits},\ }\href
  {https://doi.org/10.1103/physrevlett.62.1189} {\bibfield  {journal} {\bibinfo
   {journal} {Phys. Rev. Lett.}\ }\textbf {\bibinfo {volume} {62}},\ \bibinfo
  {pages} {1189} (\bibinfo {year} {1989})}\BibitemShut {NoStop}%
\bibitem [{\citenamefont {Binder}\ \emph {et~al.}(1990)\citenamefont {Binder},
  \citenamefont {Koch}, \citenamefont {Lindberg}, \citenamefont
  {Peyghambarian},\ and\ \citenamefont {Sch\"{a}fer}}]{Binder_Schafer_1990}%
  \BibitemOpen
  \bibfield  {author} {\bibinfo {author} {\bibfnamefont {R.}~\bibnamefont
  {Binder}}, \bibinfo {author} {\bibfnamefont {S.~W.}\ \bibnamefont {Koch}},
  \bibinfo {author} {\bibfnamefont {M.}~\bibnamefont {Lindberg}}, \bibinfo
  {author} {\bibfnamefont {N.}~\bibnamefont {Peyghambarian}},\ and\ \bibinfo
  {author} {\bibfnamefont {W.}~\bibnamefont {Sch\"{a}fer}},\ }\bibfield
  {title} {\bibinfo {title} {Ultrafast adiabatic following in semiconductors},\
  }\href {https://doi.org/10.1103/physrevlett.65.899} {\bibfield  {journal}
  {\bibinfo  {journal} {Phys. Rev. Lett.}\ }\textbf {\bibinfo {volume} {65}},\
  \bibinfo {pages} {899} (\bibinfo {year} {1990})}\BibitemShut {NoStop}%
\bibitem [{\citenamefont {Holthaus}\ and\ \citenamefont
  {Hone}(1994)}]{Holthaus_Hone_1994}%
  \BibitemOpen
  \bibfield  {author} {\bibinfo {author} {\bibfnamefont {M.}~\bibnamefont
  {Holthaus}}\ and\ \bibinfo {author} {\bibfnamefont {D.~W.}\ \bibnamefont
  {Hone}},\ }\bibfield  {title} {\bibinfo {title} {ac {Stark} effects and
  harmonic generation in periodic potentials},\ }\href
  {https://doi.org/10.1103/physrevb.49.16605} {\bibfield  {journal} {\bibinfo
  {journal} {Phys. Rev. B}\ }\textbf {\bibinfo {volume} {49}},\ \bibinfo
  {pages} {16605} (\bibinfo {year} {1994})}\BibitemShut {NoStop}%
\bibitem [{\citenamefont {Elgin}\ and\ \citenamefont
  {New}(1976)}]{Elgin_New_1976}%
  \BibitemOpen
  \bibfield  {author} {\bibinfo {author} {\bibfnamefont {J.}~\bibnamefont
  {Elgin}}\ and\ \bibinfo {author} {\bibfnamefont {G.}~\bibnamefont {New}},\
  }\bibfield  {title} {\bibinfo {title} {Semi-classical theory of two-photon
  resonant third-harmonic generation},\ }\href
  {https://doi.org/10.1016/0030-4018(76)90227-3} {\bibfield  {journal}
  {\bibinfo  {journal} {Opt. Commun.}\ }\textbf {\bibinfo {volume} {16}},\
  \bibinfo {pages} {242 } (\bibinfo {year} {1976})}\BibitemShut {NoStop}%
\bibitem [{\citenamefont {Sun}\ \emph {et~al.}(2016)\citenamefont {Sun},
  \citenamefont {Martinez},\ and\ \citenamefont {Wang}}]{Sun_Wang_2016}%
  \BibitemOpen
  \bibfield  {author} {\bibinfo {author} {\bibfnamefont {Z.}~\bibnamefont
  {Sun}}, \bibinfo {author} {\bibfnamefont {A.}~\bibnamefont {Martinez}},\ and\
  \bibinfo {author} {\bibfnamefont {F.}~\bibnamefont {Wang}},\ }\bibfield
  {title} {\bibinfo {title} {Optical modulators with {2D} layered materials},\
  }\href {https://doi.org/10.1038/nphoton.2016.15} {\bibfield  {journal}
  {\bibinfo  {journal} {Nat. Photonics}\ }\textbf {\bibinfo {volume} {10}},\
  \bibinfo {pages} {227} (\bibinfo {year} {2016})}\BibitemShut {NoStop}%
\bibitem [{\citenamefont {Cruz}\ \emph {et~al.}(1988)\citenamefont {Cruz},
  \citenamefont {Gordon}, \citenamefont {Becker}, \citenamefont {Fork},\ and\
  \citenamefont {Shank}}]{BritoCruz_Shank_1988}%
  \BibitemOpen
  \bibfield  {author} {\bibinfo {author} {\bibfnamefont {C.~B.}\ \bibnamefont
  {Cruz}}, \bibinfo {author} {\bibfnamefont {J.}~\bibnamefont {Gordon}},
  \bibinfo {author} {\bibfnamefont {P.}~\bibnamefont {Becker}}, \bibinfo
  {author} {\bibfnamefont {R.}~\bibnamefont {Fork}},\ and\ \bibinfo {author}
  {\bibfnamefont {C.}~\bibnamefont {Shank}},\ }\bibfield  {title} {\bibinfo
  {title} {Dynamics of spectral hole burning},\ }\href
  {https://doi.org/10.1109/3.122} {\bibfield  {journal} {\bibinfo  {journal}
  {{IEEE} J. Quantum Electron.}\ }\textbf {\bibinfo {volume} {24}},\ \bibinfo
  {pages} {261} (\bibinfo {year} {1988})}\BibitemShut {NoStop}%
\bibitem [{\citenamefont {Fluegel}\ \emph {et~al.}(1987)\citenamefont
  {Fluegel}, \citenamefont {Peyghambarian}, \citenamefont {Olbright},
  \citenamefont {Lindberg}, \citenamefont {Koch}, \citenamefont {Joffre},
  \citenamefont {Hulin}, \citenamefont {Migus},\ and\ \citenamefont
  {Antonetti}}]{Fleugel_Antonetti_1987}%
  \BibitemOpen
  \bibfield  {author} {\bibinfo {author} {\bibfnamefont {B.}~\bibnamefont
  {Fluegel}}, \bibinfo {author} {\bibfnamefont {N.}~\bibnamefont
  {Peyghambarian}}, \bibinfo {author} {\bibfnamefont {G.}~\bibnamefont
  {Olbright}}, \bibinfo {author} {\bibfnamefont {M.}~\bibnamefont {Lindberg}},
  \bibinfo {author} {\bibfnamefont {S.~W.}\ \bibnamefont {Koch}}, \bibinfo
  {author} {\bibfnamefont {M.}~\bibnamefont {Joffre}}, \bibinfo {author}
  {\bibfnamefont {D.}~\bibnamefont {Hulin}}, \bibinfo {author} {\bibfnamefont
  {A.}~\bibnamefont {Migus}},\ and\ \bibinfo {author} {\bibfnamefont
  {A.}~\bibnamefont {Antonetti}},\ }\bibfield  {title} {\bibinfo {title}
  {Femtosecond studies of coherent transients in semiconductors},\ }\href
  {https://doi.org/10.1103/PhysRevLett.59.2588} {\bibfield  {journal} {\bibinfo
   {journal} {Phys. Rev. Lett.}\ }\textbf {\bibinfo {volume} {59}},\ \bibinfo
  {pages} {2588} (\bibinfo {year} {1987})}\BibitemShut {NoStop}%
\bibitem [{\citenamefont {Joffre}\ \emph {et~al.}(1988)\citenamefont {Joffre},
  \citenamefont {Hulin}, \citenamefont {Migus}, \citenamefont {Antonetti},
  \citenamefont {la~Guillaume}, \citenamefont {Peyghambarian}, \citenamefont
  {Lindberg},\ and\ \citenamefont {Koch}}]{Joffre_Koch_1988}%
  \BibitemOpen
  \bibfield  {author} {\bibinfo {author} {\bibfnamefont {M.}~\bibnamefont
  {Joffre}}, \bibinfo {author} {\bibfnamefont {D.}~\bibnamefont {Hulin}},
  \bibinfo {author} {\bibfnamefont {A.}~\bibnamefont {Migus}}, \bibinfo
  {author} {\bibfnamefont {A.}~\bibnamefont {Antonetti}}, \bibinfo {author}
  {\bibfnamefont {C.~B.~a.}\ \bibnamefont {la~Guillaume}}, \bibinfo {author}
  {\bibfnamefont {N.}~\bibnamefont {Peyghambarian}}, \bibinfo {author}
  {\bibfnamefont {M.}~\bibnamefont {Lindberg}},\ and\ \bibinfo {author}
  {\bibfnamefont {S.~W.}\ \bibnamefont {Koch}},\ }\bibfield  {title} {\bibinfo
  {title} {Coherent effects in pump--probe spectroscopy of excitons},\ }\href
  {https://doi.org/10.1364/OL.13.000276} {\bibfield  {journal} {\bibinfo
  {journal} {Opt. Lett.}\ }\textbf {\bibinfo {volume} {13}},\ \bibinfo {pages}
  {276} (\bibinfo {year} {1988})}\BibitemShut {NoStop}%
\bibitem [{\citenamefont {Schmitt-Rink}\ \emph {et~al.}(1988)\citenamefont
  {Schmitt-Rink}, \citenamefont {Chemla},\ and\ \citenamefont
  {Haug}}]{SchmittRink_Haug_1988}%
  \BibitemOpen
  \bibfield  {author} {\bibinfo {author} {\bibfnamefont {S.}~\bibnamefont
  {Schmitt-Rink}}, \bibinfo {author} {\bibfnamefont {D.~S.}\ \bibnamefont
  {Chemla}},\ and\ \bibinfo {author} {\bibfnamefont {H.}~\bibnamefont {Haug}},\
  }\bibfield  {title} {\bibinfo {title} {Nonequilibrium theory of the optical
  stark effect and spectral hole burning in semiconductors},\ }\href
  {https://doi.org/10.1103/physrevb.37.941} {\bibfield  {journal} {\bibinfo
  {journal} {Phys. Rev. B}\ }\textbf {\bibinfo {volume} {37}},\ \bibinfo
  {pages} {941} (\bibinfo {year} {1988})}\BibitemShut {NoStop}%
\bibitem [{\citenamefont {Xiong}\ \emph {et~al.}(2009)\citenamefont {Xiong},
  \citenamefont {Laaser}, \citenamefont {Paoprasert}, \citenamefont {Franking},
  \citenamefont {Hamers}, \citenamefont {Gopalan},\ and\ \citenamefont
  {Zanni}}]{Xiong_Zanni_2009}%
  \BibitemOpen
  \bibfield  {author} {\bibinfo {author} {\bibfnamefont {W.}~\bibnamefont
  {Xiong}}, \bibinfo {author} {\bibfnamefont {J.~E.}\ \bibnamefont {Laaser}},
  \bibinfo {author} {\bibfnamefont {P.}~\bibnamefont {Paoprasert}}, \bibinfo
  {author} {\bibfnamefont {R.~A.}\ \bibnamefont {Franking}}, \bibinfo {author}
  {\bibfnamefont {R.~J.}\ \bibnamefont {Hamers}}, \bibinfo {author}
  {\bibfnamefont {P.}~\bibnamefont {Gopalan}},\ and\ \bibinfo {author}
  {\bibfnamefont {M.~T.}\ \bibnamefont {Zanni}},\ }\bibfield  {title} {\bibinfo
  {title} {Transient 2{D} {IR} spectroscopy of charge injection in
  dye-sensitized nanocrystalline thin films},\ }\href
  {https://doi.org/10.1021/ja908479r} {\bibfield  {journal} {\bibinfo
  {journal} {J. Am. Chem. Soc.}\ }\textbf {\bibinfo {volume} {131}},\ \bibinfo
  {pages} {18040} (\bibinfo {year} {2009})}\BibitemShut {NoStop}%
\bibitem [{\citenamefont {Dietze}\ and\ \citenamefont
  {Mathies}(2016)}]{Dietze_Mathies_2016}%
  \BibitemOpen
  \bibfield  {author} {\bibinfo {author} {\bibfnamefont {D.~R.}\ \bibnamefont
  {Dietze}}\ and\ \bibinfo {author} {\bibfnamefont {R.~A.}\ \bibnamefont
  {Mathies}},\ }\bibfield  {title} {\bibinfo {title} {Femtosecond stimulated
  {Raman} spectroscopy},\ }\href {https://doi.org/10.1002/cphc.201600104}
  {\bibfield  {journal} {\bibinfo  {journal} {{ChemPhysChem}}\ }\textbf
  {\bibinfo {volume} {17}},\ \bibinfo {pages} {1224} (\bibinfo {year}
  {2016})}\BibitemShut {NoStop}%
\bibitem [{\citenamefont {Mandal}\ \emph {et~al.}(2019)\citenamefont {Mandal},
  \citenamefont {Schultz}, \citenamefont {Wu}, \citenamefont {Coleman},
  \citenamefont {Young},\ and\ \citenamefont
  {Wasielewski}}]{Mandal_Wasielewski_2019}%
  \BibitemOpen
  \bibfield  {author} {\bibinfo {author} {\bibfnamefont {A.}~\bibnamefont
  {Mandal}}, \bibinfo {author} {\bibfnamefont {J.~D.}\ \bibnamefont {Schultz}},
  \bibinfo {author} {\bibfnamefont {Y.-L.}\ \bibnamefont {Wu}}, \bibinfo
  {author} {\bibfnamefont {A.~F.}\ \bibnamefont {Coleman}}, \bibinfo {author}
  {\bibfnamefont {R.~M.}\ \bibnamefont {Young}},\ and\ \bibinfo {author}
  {\bibfnamefont {M.~R.}\ \bibnamefont {Wasielewski}},\ }\bibfield  {title}
  {\bibinfo {title} {Transient two-dimensional electronic spectroscopy:
  Coherent dynamics at arbitrary times along the reaction coordinate},\ }\href
  {https://doi.org/10.1021/acs.jpclett.9b00826} {\bibfield  {journal} {\bibinfo
   {journal} {J. Phys. Chem. Lett.}\ }\textbf {\bibinfo {volume} {10}},\
  \bibinfo {pages} {3509} (\bibinfo {year} {2019})}\BibitemShut {NoStop}%
\bibitem [{\citenamefont {Langer}\ \emph {et~al.}(2018)\citenamefont {Langer},
  \citenamefont {Schmid}, \citenamefont {Schlauderer}, \citenamefont {Gmitra},
  \citenamefont {Fabian}, \citenamefont {Nagler}, \citenamefont {Sch\"{u}ller},
  \citenamefont {Korn}, \citenamefont {Hawkins}, \citenamefont {Steiner},
  \citenamefont {Huttner}, \citenamefont {Koch}, \citenamefont {Kira},\ and\
  \citenamefont {Huber}}]{Langer_Huber_2018}%
  \BibitemOpen
  \bibfield  {author} {\bibinfo {author} {\bibfnamefont {F.}~\bibnamefont
  {Langer}}, \bibinfo {author} {\bibfnamefont {C.~P.}\ \bibnamefont {Schmid}},
  \bibinfo {author} {\bibfnamefont {S.}~\bibnamefont {Schlauderer}}, \bibinfo
  {author} {\bibfnamefont {M.}~\bibnamefont {Gmitra}}, \bibinfo {author}
  {\bibfnamefont {J.}~\bibnamefont {Fabian}}, \bibinfo {author} {\bibfnamefont
  {P.}~\bibnamefont {Nagler}}, \bibinfo {author} {\bibfnamefont
  {C.}~\bibnamefont {Sch\"{u}ller}}, \bibinfo {author} {\bibfnamefont
  {T.}~\bibnamefont {Korn}}, \bibinfo {author} {\bibfnamefont {P.~G.}\
  \bibnamefont {Hawkins}}, \bibinfo {author} {\bibfnamefont {J.~T.}\
  \bibnamefont {Steiner}}, \bibinfo {author} {\bibfnamefont {U.}~\bibnamefont
  {Huttner}}, \bibinfo {author} {\bibfnamefont {S.~W.}\ \bibnamefont {Koch}},
  \bibinfo {author} {\bibfnamefont {M.}~\bibnamefont {Kira}},\ and\ \bibinfo
  {author} {\bibfnamefont {R.}~\bibnamefont {Huber}},\ }\bibfield  {title}
  {\bibinfo {title} {Lightwave valleytronics in a monolayer of tungsten
  diselenide},\ }\href {https://doi.org/10.1038/s41586-018-0013-6} {\bibfield
  {journal} {\bibinfo  {journal} {Nature}\ }\textbf {\bibinfo {volume} {557}},\
  \bibinfo {pages} {76} (\bibinfo {year} {2018})}\BibitemShut {NoStop}%
\bibitem [{\citenamefont {Hong}\ \emph {et~al.}(2014)\citenamefont {Hong},
  \citenamefont {Kim}, \citenamefont {Shi}, \citenamefont {Zhang},
  \citenamefont {Jin}, \citenamefont {Sun}, \citenamefont {Tongay},
  \citenamefont {Wu}, \citenamefont {Zhang},\ and\ \citenamefont
  {Wang}}]{Hong_Wang_2014}%
  \BibitemOpen
  \bibfield  {author} {\bibinfo {author} {\bibfnamefont {X.}~\bibnamefont
  {Hong}}, \bibinfo {author} {\bibfnamefont {J.}~\bibnamefont {Kim}}, \bibinfo
  {author} {\bibfnamefont {S.-F.}\ \bibnamefont {Shi}}, \bibinfo {author}
  {\bibfnamefont {Y.}~\bibnamefont {Zhang}}, \bibinfo {author} {\bibfnamefont
  {C.}~\bibnamefont {Jin}}, \bibinfo {author} {\bibfnamefont {Y.}~\bibnamefont
  {Sun}}, \bibinfo {author} {\bibfnamefont {S.}~\bibnamefont {Tongay}},
  \bibinfo {author} {\bibfnamefont {J.}~\bibnamefont {Wu}}, \bibinfo {author}
  {\bibfnamefont {Y.}~\bibnamefont {Zhang}},\ and\ \bibinfo {author}
  {\bibfnamefont {F.}~\bibnamefont {Wang}},\ }\bibfield  {title} {\bibinfo
  {title} {{Ultrafast charge transfer in atomically thin MoS$_2$/WS$_2$
  heterostructures}},\ }\href {https://doi.org/10.1038/nnano.2014.167}
  {\bibfield  {journal} {\bibinfo  {journal} {Nat. Nanotechnol.}\ }\textbf
  {\bibinfo {volume} {9}},\ \bibinfo {pages} {682} (\bibinfo {year}
  {2014})}\BibitemShut {NoStop}%
\bibitem [{\citenamefont {Malard}\ \emph {et~al.}(2013)\citenamefont {Malard},
  \citenamefont {Alencar}, \citenamefont {Barboza}, \citenamefont {Mak},\ and\
  \citenamefont {de~Paula}}]{Malard_dePaula_2013}%
  \BibitemOpen
  \bibfield  {author} {\bibinfo {author} {\bibfnamefont {L.~M.}\ \bibnamefont
  {Malard}}, \bibinfo {author} {\bibfnamefont {T.~V.}\ \bibnamefont {Alencar}},
  \bibinfo {author} {\bibfnamefont {A.~P.~M.}\ \bibnamefont {Barboza}},
  \bibinfo {author} {\bibfnamefont {K.~F.}\ \bibnamefont {Mak}},\ and\ \bibinfo
  {author} {\bibfnamefont {A.~M.}\ \bibnamefont {de~Paula}},\ }\bibfield
  {title} {\bibinfo {title} {Observation of intense second harmonic generation
  from {MoS}$_2$ atomic crystals},\ }\href
  {https://doi.org/10.1103/physrevb.87.201401} {\bibfield  {journal} {\bibinfo
  {journal} {Phys. Rev. B}\ }\textbf {\bibinfo {volume} {87}},\ \bibinfo
  {pages} {201401} (\bibinfo {year} {2013})}\BibitemShut {NoStop}%
\bibitem [{\citenamefont {Wang}\ \emph {et~al.}(2013)\citenamefont {Wang},
  \citenamefont {Chien}, \citenamefont {Kumar}, \citenamefont {Kumar},
  \citenamefont {Chiu},\ and\ \citenamefont {Zhao}}]{Wang_Zhao_2013}%
  \BibitemOpen
  \bibfield  {author} {\bibinfo {author} {\bibfnamefont {R.}~\bibnamefont
  {Wang}}, \bibinfo {author} {\bibfnamefont {H.-C.}\ \bibnamefont {Chien}},
  \bibinfo {author} {\bibfnamefont {J.}~\bibnamefont {Kumar}}, \bibinfo
  {author} {\bibfnamefont {N.}~\bibnamefont {Kumar}}, \bibinfo {author}
  {\bibfnamefont {H.-Y.}\ \bibnamefont {Chiu}},\ and\ \bibinfo {author}
  {\bibfnamefont {H.}~\bibnamefont {Zhao}},\ }\bibfield  {title} {\bibinfo
  {title} {Third-harmonic generation in ultrathin films of {MoS}$_2$},\ }\href
  {https://doi.org/10.1021/am4042542} {\bibfield  {journal} {\bibinfo
  {journal} {ACS Appl. Mater. Interfaces}\ }\textbf {\bibinfo {volume} {6}},\
  \bibinfo {pages} {314} (\bibinfo {year} {2013})}\BibitemShut {NoStop}%
\bibitem [{\citenamefont {Paradisanos}\ \emph {et~al.}(2014)\citenamefont
  {Paradisanos}, \citenamefont {Kymakis}, \citenamefont {Fotakis},
  \citenamefont {Kioseoglou},\ and\ \citenamefont
  {Stratakis}}]{Paradisanos_Stratakis_2014}%
  \BibitemOpen
  \bibfield  {author} {\bibinfo {author} {\bibfnamefont {I.}~\bibnamefont
  {Paradisanos}}, \bibinfo {author} {\bibfnamefont {E.}~\bibnamefont
  {Kymakis}}, \bibinfo {author} {\bibfnamefont {C.}~\bibnamefont {Fotakis}},
  \bibinfo {author} {\bibfnamefont {G.}~\bibnamefont {Kioseoglou}},\ and\
  \bibinfo {author} {\bibfnamefont {E.}~\bibnamefont {Stratakis}},\ }\bibfield
  {title} {\bibinfo {title} {Intense femtosecond photoexcitation of bulk and
  monolayer {MoS}$_2$},\ }\href {https://doi.org/10.1063/1.4891679} {\bibfield
  {journal} {\bibinfo  {journal} {Appl. Phys. Lett.}\ }\textbf {\bibinfo
  {volume} {105}},\ \bibinfo {pages} {041108} (\bibinfo {year}
  {2014})}\BibitemShut {NoStop}%
\bibitem [{\citenamefont {Sie}\ \emph {et~al.}(2014)\citenamefont {Sie},
  \citenamefont {McIver}, \citenamefont {Lee}, \citenamefont {Fu},
  \citenamefont {Kong},\ and\ \citenamefont {Gedik}}]{Sie_Gedik_2014}%
  \BibitemOpen
  \bibfield  {author} {\bibinfo {author} {\bibfnamefont {E.~J.}\ \bibnamefont
  {Sie}}, \bibinfo {author} {\bibfnamefont {J.~W.}\ \bibnamefont {McIver}},
  \bibinfo {author} {\bibfnamefont {Y.-H.}\ \bibnamefont {Lee}}, \bibinfo
  {author} {\bibfnamefont {L.}~\bibnamefont {Fu}}, \bibinfo {author}
  {\bibfnamefont {J.}~\bibnamefont {Kong}},\ and\ \bibinfo {author}
  {\bibfnamefont {N.}~\bibnamefont {Gedik}},\ }\bibfield  {title} {\bibinfo
  {title} {Valley-selective optical stark effect in monolayer {WS}$_2$},\
  }\href {https://doi.org/10.1038/nmat4156} {\bibfield  {journal} {\bibinfo
  {journal} {Nat. Mater.}\ }\textbf {\bibinfo {volume} {14}},\ \bibinfo {pages}
  {290} (\bibinfo {year} {2014})}\BibitemShut {NoStop}%
\bibitem [{\citenamefont {Sie}\ \emph {et~al.}(2016)\citenamefont {Sie},
  \citenamefont {Lui}, \citenamefont {Lee}, \citenamefont {Kong},\ and\
  \citenamefont {Gedik}}]{Sie_Gedik_2016}%
  \BibitemOpen
  \bibfield  {author} {\bibinfo {author} {\bibfnamefont {E.~J.}\ \bibnamefont
  {Sie}}, \bibinfo {author} {\bibfnamefont {C.~H.}\ \bibnamefont {Lui}},
  \bibinfo {author} {\bibfnamefont {Y.-H.}\ \bibnamefont {Lee}}, \bibinfo
  {author} {\bibfnamefont {J.}~\bibnamefont {Kong}},\ and\ \bibinfo {author}
  {\bibfnamefont {N.}~\bibnamefont {Gedik}},\ }\bibfield  {title} {\bibinfo
  {title} {Observation of intervalley biexcitonic optical stark effect in
  monolayer {WS}$_2$},\ }\href {https://doi.org/10.1021/acs.nanolett.6b02998}
  {\bibfield  {journal} {\bibinfo  {journal} {Nano Lett.}\ }\textbf {\bibinfo
  {volume} {16}},\ \bibinfo {pages} {7421} (\bibinfo {year}
  {2016})}\BibitemShut {NoStop}%
\bibitem [{\citenamefont {Sie}\ \emph {et~al.}(2017)\citenamefont {Sie},
  \citenamefont {Lui}, \citenamefont {Lee}, \citenamefont {Fu}, \citenamefont
  {Kong},\ and\ \citenamefont {Gedik}}]{Sie_Gedik_2017}%
  \BibitemOpen
  \bibfield  {author} {\bibinfo {author} {\bibfnamefont {E.~J.}\ \bibnamefont
  {Sie}}, \bibinfo {author} {\bibfnamefont {C.~H.}\ \bibnamefont {Lui}},
  \bibinfo {author} {\bibfnamefont {Y.-H.}\ \bibnamefont {Lee}}, \bibinfo
  {author} {\bibfnamefont {L.}~\bibnamefont {Fu}}, \bibinfo {author}
  {\bibfnamefont {J.}~\bibnamefont {Kong}},\ and\ \bibinfo {author}
  {\bibfnamefont {N.}~\bibnamefont {Gedik}},\ }\bibfield  {title} {\bibinfo
  {title} {Large, valley-exclusive {Bloch}-{Siegert} shift in monolayer
  {WS}$_2$},\ }\href {https://doi.org/10.1126/science.aal2241} {\bibfield
  {journal} {\bibinfo  {journal} {Science}\ }\textbf {\bibinfo {volume}
  {355}},\ \bibinfo {pages} {1066} (\bibinfo {year} {2017})}\BibitemShut
  {NoStop}%
\bibitem [{\citenamefont {Kim}\ \emph {et~al.}(2014)\citenamefont {Kim},
  \citenamefont {Hong}, \citenamefont {Jin}, \citenamefont {Shi}, \citenamefont
  {Chang}, \citenamefont {Chiu}, \citenamefont {Li},\ and\ \citenamefont
  {Wang}}]{Kim_Wang_2014}%
  \BibitemOpen
  \bibfield  {author} {\bibinfo {author} {\bibfnamefont {J.}~\bibnamefont
  {Kim}}, \bibinfo {author} {\bibfnamefont {X.}~\bibnamefont {Hong}}, \bibinfo
  {author} {\bibfnamefont {C.}~\bibnamefont {Jin}}, \bibinfo {author}
  {\bibfnamefont {S.-F.}\ \bibnamefont {Shi}}, \bibinfo {author} {\bibfnamefont
  {C.-Y.~S.}\ \bibnamefont {Chang}}, \bibinfo {author} {\bibfnamefont {M.-H.}\
  \bibnamefont {Chiu}}, \bibinfo {author} {\bibfnamefont {L.-J.}\ \bibnamefont
  {Li}},\ and\ \bibinfo {author} {\bibfnamefont {F.}~\bibnamefont {Wang}},\
  }\bibfield  {title} {\bibinfo {title} {Ultrafast generation of
  pseudo-magnetic field for valley excitons in {WSe}$_2$ monolayers},\ }\href
  {https://doi.org/10.1126/science.1258122} {\bibfield  {journal} {\bibinfo
  {journal} {Science}\ }\textbf {\bibinfo {volume} {346}},\ \bibinfo {pages}
  {1205} (\bibinfo {year} {2014})}\BibitemShut {NoStop}%
\bibitem [{\citenamefont {Yong}\ \emph {et~al.}(2018)\citenamefont {Yong},
  \citenamefont {Horng}, \citenamefont {Shen}, \citenamefont {Cai},
  \citenamefont {Wang}, \citenamefont {Yang}, \citenamefont {Lin},
  \citenamefont {Zhao}, \citenamefont {Watanabe}, \citenamefont {Taniguchi},
  \citenamefont {Tongay},\ and\ \citenamefont {Wang}}]{Yong_Wang_2018}%
  \BibitemOpen
  \bibfield  {author} {\bibinfo {author} {\bibfnamefont {C.-K.}\ \bibnamefont
  {Yong}}, \bibinfo {author} {\bibfnamefont {J.}~\bibnamefont {Horng}},
  \bibinfo {author} {\bibfnamefont {Y.}~\bibnamefont {Shen}}, \bibinfo {author}
  {\bibfnamefont {H.}~\bibnamefont {Cai}}, \bibinfo {author} {\bibfnamefont
  {A.}~\bibnamefont {Wang}}, \bibinfo {author} {\bibfnamefont {C.-S.}\
  \bibnamefont {Yang}}, \bibinfo {author} {\bibfnamefont {C.-K.}\ \bibnamefont
  {Lin}}, \bibinfo {author} {\bibfnamefont {S.}~\bibnamefont {Zhao}}, \bibinfo
  {author} {\bibfnamefont {K.}~\bibnamefont {Watanabe}}, \bibinfo {author}
  {\bibfnamefont {T.}~\bibnamefont {Taniguchi}}, \bibinfo {author}
  {\bibfnamefont {S.}~\bibnamefont {Tongay}},\ and\ \bibinfo {author}
  {\bibfnamefont {F.}~\bibnamefont {Wang}},\ }\bibfield  {title} {\bibinfo
  {title} {Biexcitonic optical {Stark} effects in monolayer molybdenum
  diselenide},\ }\href {https://doi.org/10.1038/s41567-018-0216-7} {\bibfield
  {journal} {\bibinfo  {journal} {Nat. Phys.}\ }\textbf {\bibinfo {volume}
  {14}},\ \bibinfo {pages} {1092} (\bibinfo {year} {2018})}\BibitemShut
  {NoStop}%
\bibitem [{\citenamefont {Yong}\ \emph {et~al.}(2019)\citenamefont {Yong},
  \citenamefont {Utama}, \citenamefont {Ong}, \citenamefont {Cao},
  \citenamefont {Regan}, \citenamefont {Horng}, \citenamefont {Shen},
  \citenamefont {Cai}, \citenamefont {Watanabe}, \citenamefont {Taniguchi},
  \citenamefont {Tongay}, \citenamefont {Deng}, \citenamefont {Zettl},
  \citenamefont {Louie},\ and\ \citenamefont {Wang}}]{Yong_Wang_2019}%
  \BibitemOpen
  \bibfield  {author} {\bibinfo {author} {\bibfnamefont {C.-K.}\ \bibnamefont
  {Yong}}, \bibinfo {author} {\bibfnamefont {M.~I.~B.}\ \bibnamefont {Utama}},
  \bibinfo {author} {\bibfnamefont {C.~S.}\ \bibnamefont {Ong}}, \bibinfo
  {author} {\bibfnamefont {T.}~\bibnamefont {Cao}}, \bibinfo {author}
  {\bibfnamefont {E.~C.}\ \bibnamefont {Regan}}, \bibinfo {author}
  {\bibfnamefont {J.}~\bibnamefont {Horng}}, \bibinfo {author} {\bibfnamefont
  {Y.}~\bibnamefont {Shen}}, \bibinfo {author} {\bibfnamefont {H.}~\bibnamefont
  {Cai}}, \bibinfo {author} {\bibfnamefont {K.}~\bibnamefont {Watanabe}},
  \bibinfo {author} {\bibfnamefont {T.}~\bibnamefont {Taniguchi}}, \bibinfo
  {author} {\bibfnamefont {S.}~\bibnamefont {Tongay}}, \bibinfo {author}
  {\bibfnamefont {H.}~\bibnamefont {Deng}}, \bibinfo {author} {\bibfnamefont
  {A.}~\bibnamefont {Zettl}}, \bibinfo {author} {\bibfnamefont {S.~G.}\
  \bibnamefont {Louie}},\ and\ \bibinfo {author} {\bibfnamefont
  {F.}~\bibnamefont {Wang}},\ }\bibfield  {title} {\bibinfo {title}
  {Valley-dependent exciton fine structure and {Autler}{\textendash}{Townes}
  doublets from {Berry} phases in monolayer {MoSe}$_2$},\ }\href
  {https://doi.org/10.1038/s41563-019-0447-8} {\bibfield  {journal} {\bibinfo
  {journal} {Nat. Mater.}\ }\textbf {\bibinfo {volume} {18}},\ \bibinfo {pages}
  {1065} (\bibinfo {year} {2019})}\BibitemShut {NoStop}%
\bibitem [{\citenamefont {LaMountain}\ \emph {et~al.}(2018)\citenamefont
  {LaMountain}, \citenamefont {Bergeron}, \citenamefont {Balla}, \citenamefont
  {Stanev}, \citenamefont {Hersam},\ and\ \citenamefont
  {Stern}}]{LaMountain_Stern_2018}%
  \BibitemOpen
  \bibfield  {author} {\bibinfo {author} {\bibfnamefont {T.}~\bibnamefont
  {LaMountain}}, \bibinfo {author} {\bibfnamefont {H.}~\bibnamefont
  {Bergeron}}, \bibinfo {author} {\bibfnamefont {I.}~\bibnamefont {Balla}},
  \bibinfo {author} {\bibfnamefont {T.~K.}\ \bibnamefont {Stanev}}, \bibinfo
  {author} {\bibfnamefont {M.~C.}\ \bibnamefont {Hersam}},\ and\ \bibinfo
  {author} {\bibfnamefont {N.~P.}\ \bibnamefont {Stern}},\ }\bibfield  {title}
  {\bibinfo {title} {Valley-selective optical {Stark} effect probed by {Kerr}
  rotation},\ }\href {https://doi.org/10.1103/physrevb.97.045307} {\bibfield
  {journal} {\bibinfo  {journal} {Phys. Rev. B}\ }\textbf {\bibinfo {volume}
  {97}},\ \bibinfo {pages} {045307} (\bibinfo {year} {2018})}\BibitemShut
  {NoStop}%
\bibitem [{\citenamefont {Cunningham}\ \emph {et~al.}(2019)\citenamefont
  {Cunningham}, \citenamefont {Hanbicki}, \citenamefont {Reinecke},
  \citenamefont {McCreary},\ and\ \citenamefont
  {Jonker}}]{Cunningham_Jonker_2019}%
  \BibitemOpen
  \bibfield  {author} {\bibinfo {author} {\bibfnamefont {P.~D.}\ \bibnamefont
  {Cunningham}}, \bibinfo {author} {\bibfnamefont {A.~T.}\ \bibnamefont
  {Hanbicki}}, \bibinfo {author} {\bibfnamefont {T.~L.}\ \bibnamefont
  {Reinecke}}, \bibinfo {author} {\bibfnamefont {K.~M.}\ \bibnamefont
  {McCreary}},\ and\ \bibinfo {author} {\bibfnamefont {B.~T.}\ \bibnamefont
  {Jonker}},\ }\bibfield  {title} {\bibinfo {title} {Resonant optical {Stark}
  effect in monolayer {WS}$_2$},\ }\href
  {https://doi.org/10.1038/s41467-019-13501-x} {\bibfield  {journal} {\bibinfo
  {journal} {Nat. Commun.}\ }\textbf {\bibinfo {volume} {10}},\ \bibinfo
  {pages} {5539} (\bibinfo {year} {2019})}\BibitemShut {NoStop}%
\bibitem [{Note1()}]{Note1}%
  \BibitemOpen
  \bibinfo {note} {See Supplemental Material at [URL will be inserted by
  publisher] for Materials and Methods, a unified perturbation theory of the
  OSE for weak field and harmonic probes, additional numerical simulations of
  pump-THG-probe, comparison of pump-SHG-probe and our OSE theory,
  demonstration of probe-induced OSE, and additional pump-THG-probe
  measurements on various WS\protect \textsubscript {2}
  morphologies.}\BibitemShut {Stop}%
\bibitem [{\citenamefont {Zhao}\ and\ \citenamefont
  {Jin}(2019)}]{Zhao_Jin_2019}%
  \BibitemOpen
  \bibfield  {author} {\bibinfo {author} {\bibfnamefont {Y.}~\bibnamefont
  {Zhao}}\ and\ \bibinfo {author} {\bibfnamefont {S.}~\bibnamefont {Jin}},\
  }\bibfield  {title} {\bibinfo {title} {Controllable water vapor assisted
  chemical vapor transport synthesis of {WS}$_2${\textendash}{MoS}$_2$
  heterostructure},\ }\href {https://doi.org/10.1021/acsmaterialslett.9b00415}
  {\bibfield  {journal} {\bibinfo  {journal} {{ACS} Materials Lett.}\ }\textbf
  {\bibinfo {volume} {2}},\ \bibinfo {pages} {42} (\bibinfo {year}
  {2019})}\BibitemShut {NoStop}%
\bibitem [{\citenamefont {Shearer}\ \emph {et~al.}(2017)\citenamefont
  {Shearer}, \citenamefont {Samad}, \citenamefont {Zhang}, \citenamefont
  {Zhao}, \citenamefont {Puretzky}, \citenamefont {Eliceiri}, \citenamefont
  {Wright}, \citenamefont {Hamers},\ and\ \citenamefont
  {Jin}}]{Shearer_Jin_2017}%
  \BibitemOpen
  \bibfield  {author} {\bibinfo {author} {\bibfnamefont {M.~J.}\ \bibnamefont
  {Shearer}}, \bibinfo {author} {\bibfnamefont {L.}~\bibnamefont {Samad}},
  \bibinfo {author} {\bibfnamefont {Y.}~\bibnamefont {Zhang}}, \bibinfo
  {author} {\bibfnamefont {Y.}~\bibnamefont {Zhao}}, \bibinfo {author}
  {\bibfnamefont {A.}~\bibnamefont {Puretzky}}, \bibinfo {author}
  {\bibfnamefont {K.~W.}\ \bibnamefont {Eliceiri}}, \bibinfo {author}
  {\bibfnamefont {J.~C.}\ \bibnamefont {Wright}}, \bibinfo {author}
  {\bibfnamefont {R.~J.}\ \bibnamefont {Hamers}},\ and\ \bibinfo {author}
  {\bibfnamefont {S.}~\bibnamefont {Jin}},\ }\bibfield  {title} {\bibinfo
  {title} {Complex and noncentrosymmetric stacking of layered metal
  dichalcogenide materials created by screw dislocations},\ }\href
  {https://doi.org/10.1021/jacs.6b12559} {\bibfield  {journal} {\bibinfo
  {journal} {J. Am. Chem. Soc.}\ }\textbf {\bibinfo {volume} {139}},\ \bibinfo
  {pages} {3496} (\bibinfo {year} {2017})}\BibitemShut {NoStop}%
\bibitem [{\citenamefont {Fan}\ \emph {et~al.}(2017)\citenamefont {Fan},
  \citenamefont {Jiang}, \citenamefont {Zhuang}, \citenamefont {Liu},
  \citenamefont {Xu}, \citenamefont {Zheng}, \citenamefont {Fan}, \citenamefont
  {Li}, \citenamefont {Wu}, \citenamefont {Zhu}, \citenamefont {Zhang},
  \citenamefont {Zhou}, \citenamefont {Hu}, \citenamefont {Wang}, \citenamefont
  {Sun}, \citenamefont {Duan},\ and\ \citenamefont {Pan}}]{Fan_Pan_2017}%
  \BibitemOpen
  \bibfield  {author} {\bibinfo {author} {\bibfnamefont {X.}~\bibnamefont
  {Fan}}, \bibinfo {author} {\bibfnamefont {Y.}~\bibnamefont {Jiang}}, \bibinfo
  {author} {\bibfnamefont {X.}~\bibnamefont {Zhuang}}, \bibinfo {author}
  {\bibfnamefont {H.}~\bibnamefont {Liu}}, \bibinfo {author} {\bibfnamefont
  {T.}~\bibnamefont {Xu}}, \bibinfo {author} {\bibfnamefont {W.}~\bibnamefont
  {Zheng}}, \bibinfo {author} {\bibfnamefont {P.}~\bibnamefont {Fan}}, \bibinfo
  {author} {\bibfnamefont {H.}~\bibnamefont {Li}}, \bibinfo {author}
  {\bibfnamefont {X.}~\bibnamefont {Wu}}, \bibinfo {author} {\bibfnamefont
  {X.}~\bibnamefont {Zhu}}, \bibinfo {author} {\bibfnamefont {Q.}~\bibnamefont
  {Zhang}}, \bibinfo {author} {\bibfnamefont {H.}~\bibnamefont {Zhou}},
  \bibinfo {author} {\bibfnamefont {W.}~\bibnamefont {Hu}}, \bibinfo {author}
  {\bibfnamefont {X.}~\bibnamefont {Wang}}, \bibinfo {author} {\bibfnamefont
  {L.}~\bibnamefont {Sun}}, \bibinfo {author} {\bibfnamefont {X.}~\bibnamefont
  {Duan}},\ and\ \bibinfo {author} {\bibfnamefont {A.}~\bibnamefont {Pan}},\
  }\bibfield  {title} {\bibinfo {title} {Broken symmetry induced strong
  nonlinear optical effects in spiral {WS}$_2$ nanosheets},\ }\href
  {https://doi.org/10.1021/acsnano.7b01457} {\bibfield  {journal} {\bibinfo
  {journal} {{ACS} Nano}\ }\textbf {\bibinfo {volume} {11}},\ \bibinfo {pages}
  {4892} (\bibinfo {year} {2017})}\BibitemShut {NoStop}%
\bibitem [{\citenamefont {Fan}\ \emph {et~al.}(2018)\citenamefont {Fan},
  \citenamefont {Zhao}, \citenamefont {Zheng}, \citenamefont {Li},
  \citenamefont {Wu}, \citenamefont {Hu}, \citenamefont {Zhang}, \citenamefont
  {Zhu}, \citenamefont {Zhang}, \citenamefont {Wang}, \citenamefont {Yang},
  \citenamefont {Chen}, \citenamefont {Jin},\ and\ \citenamefont
  {Pan}}]{Fan_Pan_2018}%
  \BibitemOpen
  \bibfield  {author} {\bibinfo {author} {\bibfnamefont {X.}~\bibnamefont
  {Fan}}, \bibinfo {author} {\bibfnamefont {Y.}~\bibnamefont {Zhao}}, \bibinfo
  {author} {\bibfnamefont {W.}~\bibnamefont {Zheng}}, \bibinfo {author}
  {\bibfnamefont {H.}~\bibnamefont {Li}}, \bibinfo {author} {\bibfnamefont
  {X.}~\bibnamefont {Wu}}, \bibinfo {author} {\bibfnamefont {X.}~\bibnamefont
  {Hu}}, \bibinfo {author} {\bibfnamefont {X.}~\bibnamefont {Zhang}}, \bibinfo
  {author} {\bibfnamefont {X.}~\bibnamefont {Zhu}}, \bibinfo {author}
  {\bibfnamefont {Q.}~\bibnamefont {Zhang}}, \bibinfo {author} {\bibfnamefont
  {X.}~\bibnamefont {Wang}}, \bibinfo {author} {\bibfnamefont {B.}~\bibnamefont
  {Yang}}, \bibinfo {author} {\bibfnamefont {J.}~\bibnamefont {Chen}}, \bibinfo
  {author} {\bibfnamefont {S.}~\bibnamefont {Jin}},\ and\ \bibinfo {author}
  {\bibfnamefont {A.}~\bibnamefont {Pan}},\ }\bibfield  {title} {\bibinfo
  {title} {Controllable growth and formation mechanisms of dislocated {WS}$_2$
  spirals},\ }\href {https://doi.org/10.1021/acs.nanolett.8b01210} {\bibfield
  {journal} {\bibinfo  {journal} {Nano Lett.}\ }\textbf {\bibinfo {volume}
  {18}},\ \bibinfo {pages} {3885} (\bibinfo {year} {2018})}\BibitemShut
  {NoStop}%
\bibitem [{\citenamefont {Yang}\ \emph {et~al.}(2016)\citenamefont {Yang},
  \citenamefont {Yang}, \citenamefont {Zhu}, \citenamefont {Johnson},
  \citenamefont {Berry}, \citenamefont {van~de Lagemaat},\ and\ \citenamefont
  {Beard}}]{Yang_Beard_2016}%
  \BibitemOpen
  \bibfield  {author} {\bibinfo {author} {\bibfnamefont {Y.}~\bibnamefont
  {Yang}}, \bibinfo {author} {\bibfnamefont {M.}~\bibnamefont {Yang}}, \bibinfo
  {author} {\bibfnamefont {K.}~\bibnamefont {Zhu}}, \bibinfo {author}
  {\bibfnamefont {J.~C.}\ \bibnamefont {Johnson}}, \bibinfo {author}
  {\bibfnamefont {J.~J.}\ \bibnamefont {Berry}}, \bibinfo {author}
  {\bibfnamefont {J.}~\bibnamefont {van~de Lagemaat}},\ and\ \bibinfo {author}
  {\bibfnamefont {M.~C.}\ \bibnamefont {Beard}},\ }\bibfield  {title} {\bibinfo
  {title} {Large polarization-dependent exciton optical {Stark} effect in lead
  iodide perovskites},\ }\href {https://doi.org/10.1038/ncomms12613} {\bibfield
   {journal} {\bibinfo  {journal} {Nat.Commun.}\ }\textbf {\bibinfo {volume}
  {7}},\ \bibinfo {pages} {12613} (\bibinfo {year} {2016})}\BibitemShut
  {NoStop}%
\bibitem [{\citenamefont {Proppe}\ \emph {et~al.}(2020)\citenamefont {Proppe},
  \citenamefont {Walters}, \citenamefont {Alsalloum}, \citenamefont
  {Zhumekenov}, \citenamefont {Mosconi}, \citenamefont {Kelley}, \citenamefont
  {Angelis}, \citenamefont {Adamska}, \citenamefont {Umari}, \citenamefont
  {Bakr},\ and\ \citenamefont {Sargent}}]{Proppe_Sargent_2020}%
  \BibitemOpen
  \bibfield  {author} {\bibinfo {author} {\bibfnamefont {A.~H.}\ \bibnamefont
  {Proppe}}, \bibinfo {author} {\bibfnamefont {G.~W.}\ \bibnamefont {Walters}},
  \bibinfo {author} {\bibfnamefont {A.~Y.}\ \bibnamefont {Alsalloum}}, \bibinfo
  {author} {\bibfnamefont {A.~A.}\ \bibnamefont {Zhumekenov}}, \bibinfo
  {author} {\bibfnamefont {E.}~\bibnamefont {Mosconi}}, \bibinfo {author}
  {\bibfnamefont {S.~O.}\ \bibnamefont {Kelley}}, \bibinfo {author}
  {\bibfnamefont {F.~D.}\ \bibnamefont {Angelis}}, \bibinfo {author}
  {\bibfnamefont {L.}~\bibnamefont {Adamska}}, \bibinfo {author} {\bibfnamefont
  {P.}~\bibnamefont {Umari}}, \bibinfo {author} {\bibfnamefont {O.~M.}\
  \bibnamefont {Bakr}},\ and\ \bibinfo {author} {\bibfnamefont {E.~H.}\
  \bibnamefont {Sargent}},\ }\bibfield  {title} {\bibinfo {title} {Transition
  dipole moments of n = 1, 2, and 3 perovskite quantum wells from the optical
  {Stark} effect and many-body perturbation theory},\ }\href
  {https://doi.org/10.1021/acs.jpclett.9b03349} {\bibfield  {journal} {\bibinfo
   {journal} {J. Phys. Chem. Lett.}\ }\textbf {\bibinfo {volume} {11}},\
  \bibinfo {pages} {716} (\bibinfo {year} {2020})}\BibitemShut {NoStop}%
\bibitem [{\citenamefont {Mukamel}(1999)}]{Mukamel_1999}%
  \BibitemOpen
  \bibfield  {author} {\bibinfo {author} {\bibfnamefont {S.}~\bibnamefont
  {Mukamel}},\ }\href@noop {} {\emph {\bibinfo {title} {Principles of Nonlinear
  Optical Spectroscopy}}}\ (\bibinfo  {publisher} {Oxford University Press},\
  \bibinfo {year} {1999})\BibitemShut {NoStop}%
\bibitem [{\citenamefont {Lee}\ and\ \citenamefont
  {Albrecht}(1985)}]{Lee_Albrecht_1985}%
  \BibitemOpen
  \bibfield  {author} {\bibinfo {author} {\bibfnamefont {D.}~\bibnamefont
  {Lee}}\ and\ \bibinfo {author} {\bibfnamefont {A.~C.}\ \bibnamefont
  {Albrecht}},\ }\bibfield  {title} {\bibinfo {title} {{A Unified View of
  Raman, Resonance Raman, and Fluorescence Spectroscopy (and their Analogues in
  Two-Photon Absorption)}},\ }in\ \href@noop {} {\emph {\bibinfo {booktitle}
  {Advances in Infrared and Raman Spectroscopies}}},\ Vol.~\bibinfo {volume}
  {12}\ (\bibinfo {year} {1985})\ pp.\ \bibinfo {pages} {179--213}\BibitemShut
  {NoStop}%
\bibitem [{\citenamefont {Lin}\ \emph {et~al.}(2019)\citenamefont {Lin},
  \citenamefont {Bange},\ and\ \citenamefont {Lupton}}]{Lin_Lupton_2019}%
  \BibitemOpen
  \bibfield  {author} {\bibinfo {author} {\bibfnamefont {K.-Q.}\ \bibnamefont
  {Lin}}, \bibinfo {author} {\bibfnamefont {S.}~\bibnamefont {Bange}},\ and\
  \bibinfo {author} {\bibfnamefont {J.~M.}\ \bibnamefont {Lupton}},\ }\bibfield
   {title} {\bibinfo {title} {Quantum interference in second-harmonic
  generation from monolayer {WSe}$_2$},\ }\href
  {https://doi.org/10.1038/s41567-018-0384-5} {\bibfield  {journal} {\bibinfo
  {journal} {Nat. Phys.}\ }\textbf {\bibinfo {volume} {15}},\ \bibinfo {pages}
  {242} (\bibinfo {year} {2019})}\BibitemShut {NoStop}%
\bibitem [{\citenamefont {Wang}\ \emph {et~al.}(2019)\citenamefont {Wang},
  \citenamefont {Muniz}, \citenamefont {Sipe},\ and\ \citenamefont
  {Cundiff}}]{Wang_Cundiff_2019}%
  \BibitemOpen
  \bibfield  {author} {\bibinfo {author} {\bibfnamefont {K.}~\bibnamefont
  {Wang}}, \bibinfo {author} {\bibfnamefont {R.~A.}\ \bibnamefont {Muniz}},
  \bibinfo {author} {\bibfnamefont {J.~E.}\ \bibnamefont {Sipe}},\ and\
  \bibinfo {author} {\bibfnamefont {S.~T.}\ \bibnamefont {Cundiff}},\
  }\bibfield  {title} {\bibinfo {title} {Quantum interference control of
  photocurrents in semiconductors by nonlinear optical absorption processes},\
  }\href {https://doi.org/10.1103/physrevlett.123.067402} {\bibfield  {journal}
  {\bibinfo  {journal} {Phys. Rev. Lett.}\ }\textbf {\bibinfo {volume} {123}},\
  \bibinfo {pages} {067402} (\bibinfo {year} {2019})}\BibitemShut {NoStop}%
\bibitem [{\citenamefont {Mahon}\ \emph {et~al.}(2019)\citenamefont {Mahon},
  \citenamefont {Muniz},\ and\ \citenamefont {Sipe}}]{Mahon_Sipe_2019}%
  \BibitemOpen
  \bibfield  {author} {\bibinfo {author} {\bibfnamefont {P.~T.}\ \bibnamefont
  {Mahon}}, \bibinfo {author} {\bibfnamefont {R.~A.}\ \bibnamefont {Muniz}},\
  and\ \bibinfo {author} {\bibfnamefont {J.~E.}\ \bibnamefont {Sipe}},\
  }\bibfield  {title} {\bibinfo {title} {Quantum interference control of
  localized carrier distributions in the {Brillouin} zone},\ }\href
  {https://doi.org/10.1103/physrevb.100.075203} {\bibfield  {journal} {\bibinfo
   {journal} {Phys. Rev. B}\ }\textbf {\bibinfo {volume} {100}},\ \bibinfo
  {pages} {075203} (\bibinfo {year} {2019})}\BibitemShut {NoStop}%
\bibitem [{\citenamefont {Muniz}\ \emph {et~al.}(2019)\citenamefont {Muniz},
  \citenamefont {Salazar}, \citenamefont {Wang}, \citenamefont {Cundiff},\ and\
  \citenamefont {Sipe}}]{Muniz_Sipe_2019}%
  \BibitemOpen
  \bibfield  {author} {\bibinfo {author} {\bibfnamefont {R.~A.}\ \bibnamefont
  {Muniz}}, \bibinfo {author} {\bibfnamefont {C.}~\bibnamefont {Salazar}},
  \bibinfo {author} {\bibfnamefont {K.}~\bibnamefont {Wang}}, \bibinfo {author}
  {\bibfnamefont {S.~T.}\ \bibnamefont {Cundiff}},\ and\ \bibinfo {author}
  {\bibfnamefont {J.~E.}\ \bibnamefont {Sipe}},\ }\bibfield  {title} {\bibinfo
  {title} {Quantum interference control of carriers and currents in zinc blende
  semiconductors based on nonlinear absorption processes},\ }\href
  {https://doi.org/10.1103/physrevb.100.075202} {\bibfield  {journal} {\bibinfo
   {journal} {Phys. Rev. B}\ }\textbf {\bibinfo {volume} {100}},\ \bibinfo
  {pages} {075202} (\bibinfo {year} {2019})}\BibitemShut {NoStop}%
\bibitem [{\citenamefont {Kohler}\ \emph {et~al.}(2017)\citenamefont {Kohler},
  \citenamefont {Thompson},\ and\ \citenamefont {Wright}}]{Kohler_Wright_2017}%
  \BibitemOpen
  \bibfield  {author} {\bibinfo {author} {\bibfnamefont {D.~D.}\ \bibnamefont
  {Kohler}}, \bibinfo {author} {\bibfnamefont {B.~J.}\ \bibnamefont
  {Thompson}},\ and\ \bibinfo {author} {\bibfnamefont {J.~C.}\ \bibnamefont
  {Wright}},\ }\bibfield  {title} {\bibinfo {title} {Frequency-domain coherent
  multidimensional spectroscopy when dephasing rivals pulsewidth: Disentangling
  material and instrument response},\ }\href
  {https://doi.org/10.1063/1.4986069} {\bibfield  {journal} {\bibinfo
  {journal} {J. Chem. Phys.}\ }\textbf {\bibinfo {volume} {147}},\ \bibinfo
  {pages} {084202} (\bibinfo {year} {2017})}\BibitemShut {NoStop}%
\bibitem [{\citenamefont {Gelin}\ \emph {et~al.}(2005)\citenamefont {Gelin},
  \citenamefont {Egorova},\ and\ \citenamefont {Domcke}}]{Gelin_Domcke_2005}%
  \BibitemOpen
  \bibfield  {author} {\bibinfo {author} {\bibfnamefont {M.~F.}\ \bibnamefont
  {Gelin}}, \bibinfo {author} {\bibfnamefont {D.}~\bibnamefont {Egorova}},\
  and\ \bibinfo {author} {\bibfnamefont {W.}~\bibnamefont {Domcke}},\
  }\bibfield  {title} {\bibinfo {title} {Efficient method for the calculation
  of time- and frequency-resolved four-wave mixing signals and its application
  to photon-echo spectroscopy},\ }\href {https://doi.org/10.1063/1.2062188}
  {\bibfield  {journal} {\bibinfo  {journal} {J. Chem. Phys.}\ }\textbf
  {\bibinfo {volume} {123}},\ \bibinfo {pages} {164112} (\bibinfo {year}
  {2005})}\BibitemShut {NoStop}%
\bibitem [{\citenamefont {Thompson}\ \emph {et~al.}(2020)\citenamefont
  {Thompson}, \citenamefont {Sunden},\ and\ \citenamefont
  {Kohler}}]{WrightSim}%
  \BibitemOpen
  \bibfield  {author} {\bibinfo {author} {\bibfnamefont {B.}~\bibnamefont
  {Thompson}}, \bibinfo {author} {\bibfnamefont {K.}~\bibnamefont {Sunden}},\
  and\ \bibinfo {author} {\bibfnamefont {D.}~\bibnamefont {Kohler}},\ }\href
  {https://doi.org/10.5281/zenodo.3774515} {\bibinfo {title} {{WrightSim}}}
  (\bibinfo {year} {2020})\BibitemShut {NoStop}%
\bibitem [{\citenamefont {Blum}\ \emph {et~al.}(1993)\citenamefont {Blum},
  \citenamefont {Harshman}, \citenamefont {Gustafson},\ and\ \citenamefont
  {Kelley}}]{Blum_Kelley_1993}%
  \BibitemOpen
  \bibfield  {author} {\bibinfo {author} {\bibfnamefont {O.}~\bibnamefont
  {Blum}}, \bibinfo {author} {\bibfnamefont {P.}~\bibnamefont {Harshman}},
  \bibinfo {author} {\bibfnamefont {T.~K.}\ \bibnamefont {Gustafson}},\ and\
  \bibinfo {author} {\bibfnamefont {P.~L.}\ \bibnamefont {Kelley}},\ }\bibfield
   {title} {\bibinfo {title} {Application of radiative renormalization to
  strong-field resonant nonlinear optical interactions},\ }\href
  {https://doi.org/10.1103/physreva.47.5165} {\bibfield  {journal} {\bibinfo
  {journal} {Phys. Rev. A}\ }\textbf {\bibinfo {volume} {47}},\ \bibinfo
  {pages} {5165} (\bibinfo {year} {1993})}\BibitemShut {NoStop}%
\bibitem [{\citenamefont {Ye}\ \emph {et~al.}(2014)\citenamefont {Ye},
  \citenamefont {Cao}, \citenamefont {O'Brien}, \citenamefont {Zhu},
  \citenamefont {Yin}, \citenamefont {Wang}, \citenamefont {Louie},\ and\
  \citenamefont {Zhang}}]{Ye_Zhang_2014}%
  \BibitemOpen
  \bibfield  {author} {\bibinfo {author} {\bibfnamefont {Z.}~\bibnamefont
  {Ye}}, \bibinfo {author} {\bibfnamefont {T.}~\bibnamefont {Cao}}, \bibinfo
  {author} {\bibfnamefont {K.}~\bibnamefont {O'Brien}}, \bibinfo {author}
  {\bibfnamefont {H.}~\bibnamefont {Zhu}}, \bibinfo {author} {\bibfnamefont
  {X.}~\bibnamefont {Yin}}, \bibinfo {author} {\bibfnamefont {Y.}~\bibnamefont
  {Wang}}, \bibinfo {author} {\bibfnamefont {S.~G.}\ \bibnamefont {Louie}},\
  and\ \bibinfo {author} {\bibfnamefont {X.}~\bibnamefont {Zhang}},\ }\bibfield
   {title} {\bibinfo {title} {Probing excitonic dark states in single-layer
  tungsten disulphide},\ }\href {https://doi.org/10.1038/nature13734}
  {\bibfield  {journal} {\bibinfo  {journal} {Nature}\ }\textbf {\bibinfo
  {volume} {513}},\ \bibinfo {pages} {214} (\bibinfo {year}
  {2014})}\BibitemShut {NoStop}%
\bibitem [{\citenamefont {Liu}\ \emph {et~al.}(2016)\citenamefont {Liu},
  \citenamefont {Li}, \citenamefont {You}, \citenamefont {Ghimire},
  \citenamefont {Heinz},\ and\ \citenamefont {Reis}}]{Liu_Reis_2016}%
  \BibitemOpen
  \bibfield  {author} {\bibinfo {author} {\bibfnamefont {H.}~\bibnamefont
  {Liu}}, \bibinfo {author} {\bibfnamefont {Y.}~\bibnamefont {Li}}, \bibinfo
  {author} {\bibfnamefont {Y.~S.}\ \bibnamefont {You}}, \bibinfo {author}
  {\bibfnamefont {S.}~\bibnamefont {Ghimire}}, \bibinfo {author} {\bibfnamefont
  {T.~F.}\ \bibnamefont {Heinz}},\ and\ \bibinfo {author} {\bibfnamefont
  {D.~A.}\ \bibnamefont {Reis}},\ }\bibfield  {title} {\bibinfo {title}
  {High-harmonic generation from an atomically thin semiconductor},\ }\href
  {https://doi.org/10.1038/nphys3946} {\bibfield  {journal} {\bibinfo
  {journal} {Nat. Phys.}\ }\textbf {\bibinfo {volume} {13}},\ \bibinfo {pages}
  {262} (\bibinfo {year} {2016})}\BibitemShut {NoStop}%
\bibitem [{\citenamefont {Ghimire}\ and\ \citenamefont
  {Reis}(2019)}]{Ghimire_Reis_2019}%
  \BibitemOpen
  \bibfield  {author} {\bibinfo {author} {\bibfnamefont {S.}~\bibnamefont
  {Ghimire}}\ and\ \bibinfo {author} {\bibfnamefont {D.~A.}\ \bibnamefont
  {Reis}},\ }\bibfield  {title} {\bibinfo {title} {{High-harmonic generation
  from solids}},\ }\href {https://doi.org/10.1038/s41567-018-0315-5} {\bibfield
   {journal} {\bibinfo  {journal} {Nat. Phys.}\ }\textbf {\bibinfo {volume}
  {15}},\ \bibinfo {pages} {10} (\bibinfo {year} {2019})}\BibitemShut {NoStop}%
\end{thebibliography}%

\end{document}